\documentclass[reprint,a4paper,prb,superscriptaddress,citeautoscript,floatfix,showkeys]{revtex4-2}

\pdfoutput=1						

\usepackage{graphicx}
\usepackage{amsmath}
\newcommand{\parallelsum}{\mathbin{\!/\mkern-5mu/\!}}
\usepackage{verbatim} 
\begin{document}

\title{Observation of the anomalous Hall effect in a layered polar semiconductor}

\author{Seo-Jin Kim}
\affiliation{Max Planck Institute for Chemical Physics of Solids, 01187 Dresden, Germany}

\author{Jihang Zhu}
\affiliation{Max Planck Institute for the Physics of Complex Systems, 01187 Dresden, Germany}

\author{Mario M. Piva}
\affiliation{Max Planck Institute for Chemical Physics of Solids, 01187 Dresden, Germany}

\author{Marcus Schmidt}
\affiliation{Max Planck Institute for Chemical Physics of Solids, 01187 Dresden, Germany}

\author{Dorsa Fartab}
\affiliation{Max Planck Institute for Chemical Physics of Solids, 01187 Dresden, Germany}

\author{Andrew P. Mackenzie}
\email{andy.mackenzie@cpfs.mpg.de}
\affiliation{Max Planck Institute for Chemical Physics of Solids, 01187 Dresden, Germany}
\affiliation{Scottish Universities Physics Alliance, School of Physics and Astronomy,University of St Andrews, St Andrews KY16 9SS, United Kingdom}

\author{Michael Baenitz}
\affiliation{Max Planck Institute for Chemical Physics of Solids, 01187 Dresden, Germany}

\author{Michael Nicklas}
\affiliation{Max Planck Institute for Chemical Physics of Solids, 01187 Dresden, Germany}

\author{Helge Rosner}
\affiliation{Max Planck Institute for Chemical Physics of Solids, 01187 Dresden, Germany}

\author{Ashley M. Cook}
\affiliation{Max Planck Institute for Chemical Physics of Solids, 01187 Dresden, Germany}
\affiliation{Max Planck Institute for the Physics of Complex Systems, 01187 Dresden, Germany}

\author{Rafael González-Hernández}
\affiliation{Institut für Physik, Johannes Gutenberg Universität Mainz, 55128 Mainz, Germany}
\affiliation{Grupo de Investigación en Física Aplicada, Departamento de Física, Universidad del Norte, Barranquilla, Colombia}

\author{Libor Šmejkal}
\email{lsmejkal@uni-mainz.de}
\affiliation{Institut für Physik, Johannes Gutenberg Universität Mainz, 55128 Mainz, Germany}
\affiliation{Institute of Physics, Czech Academy of Sciences, Cukrovarnická 10, 162 00 Praha 6, Czech Republic}

\author{Haijing Zhang}
\email{haijing.zhang@cpfs.mpg.de}
\affiliation{Max Planck Institute for Chemical Physics of Solids, 01187 Dresden, Germany}

\begin{abstract}

Progress in magnetoelectric materials is hindered by apparently contradictory requirements for time-reversal symmetry broken and polar ferroelectric electronic structure in common ferromagnets and antiferromagnets.  Alternative routes could be provided by recent discoveries of a time-reversal symmetry breaking anomalous Hall effect in noncollinear magnets and altermagnets, but hitherto reported bulk materials are not polar. Here, we report the observation of a spontaneous anomalous Hall effect in doped AgCrSe$_2$, a layered polar semiconductor with an antiferromagnetic coupling between Cr spins in adjacent layers. The anomalous Hall resistivity 3 $\mu\Omega \, \textnormal {cm}$ is comparable to the largest observed in compensated magnetic systems to date, and is rapidly switched off when the angle of an applied magnetic field is rotated to $\sim80^\circ$ from the crystalline $c$-axis. Our ionic gating experiments show that the anomalous Hall conductivity magnitude can be enhanced by modulating the $p$-type carrier density. We also present theoretical results that suggest the anomalous Hall effect is driven by Berry curvature due to noncollinear antiferromagnetic correlations among Cr spins, which are consistent with the previously suggested magnetic ordering in AgCrSe$_2$. Our results open the possibility to study the interplay of magnetic and ferroelectric-like responses in this fascinating class of materials. 
\end{abstract}

\keywords{anomalous Hall effect, magnetism, polar structure, ionic gating, Berry curvature}

\maketitle

The anomalous Hall effect (AHE), in which electrons acquire a transverse velocity relative to an applied electric field in the absence of a magnetic field, is one of the most fundamental phenomena in condensed matter physics~\cite{RevModPhys.82.1539,vsmejkal2022anomalous}. Developments in theory based on Berry-phase concepts have provided a comprehensive framework for understanding the AHE, which may occur not only in ferromagnets, but also more generally in magnetically compensated systems with broken time-reversal symmetry (TRS) in their momentum space electronic structure, such as noncollinear kagome magnets and altermagnets~\cite{RevModPhys.82.1959,vsmejkal2020crystal,10.1063/5.0005017,Mazin2021,guin20212d,PhysRevX.12.040501,PhysRevX.12.031042,PhysRevLett.130.036702,feng2022anomalous}. 

Among many predictions and observations in this rapidly-moving field, the case of an AHE associated with a polar structure stands out. 
Polar materials can exhibit ferroelectricity and spin-orbit interaction induced spin polarization~\cite{1984JETPL,manchon2015new}. When TRS breaking in electronic structure coexists in such a system, the interplay between magnetic order and polarity creates a promising platform for the development of spintronic and magnetoelectric devices with rich functionality~\cite{avci2015unidirectional,fiebig2016evolution,ideue2017bulk}. Although there are reports that combine magnetic order and polarity at the interfaces or in heterostructures~\cite{avci2015unidirectional,brehin2023coexistence}, it is a formidable task to realize the coexistence in a single bulk material platform.

In terms of material physics, a minimal ingredient for such an observation is a system with a crystal and magnetic structure that allows for the coexistence of a polar vector, $\boldsymbol{P}$, and a Hall pseudovector, $\boldsymbol{\sigma} = (\sigma_{yz},\sigma_{zx},\sigma_{xy})$, where the components represent anomalous Hall conductivities~\cite{vsmejkal2022anomalous,landau1995electrodynamics}. Particularly interesting classes with polar interfaces intrinsic to their layered structure include layered triangular-lattice delafossite systems ~\cite{sunko2017maximal}.
Previous work has reported ferroelectricity in AgCrS$_2$ and CuCrS$_2$---layered magnetic semiconductors that have a similar triangular-lattice framework~\cite{PhysRevLett.101.067204, Singh2009, xiang2022}. This suggests that a large effect may be visible in related layered polar materials, and an outstanding question arises whether this class of crystals can also exhibit an AHE.

Here, we report that one such a doped semiconductor, AgCrSe$_2$, fulfills the symmetry requirements of polar crystal structure, and shows a spontaneous AHE. We demonstrate that it is an intrinsic AHE by comparing its evolution to that of the magnetization, and perform a consistency check by showing that the magnitude of the anomalous Hall conductivity is temperature- and scattering-independent below 50 K.  We also report and discuss the observation of a rather pronounced plateau in the Hall resistivity as a function of the angle of the applied field relative to the crystalline $c$-axis, and show that the magnitude of the observed AHE is tunable by the application of an ionic gate. \\

\begin{figure}
\centering
\includegraphics[width=1\linewidth]{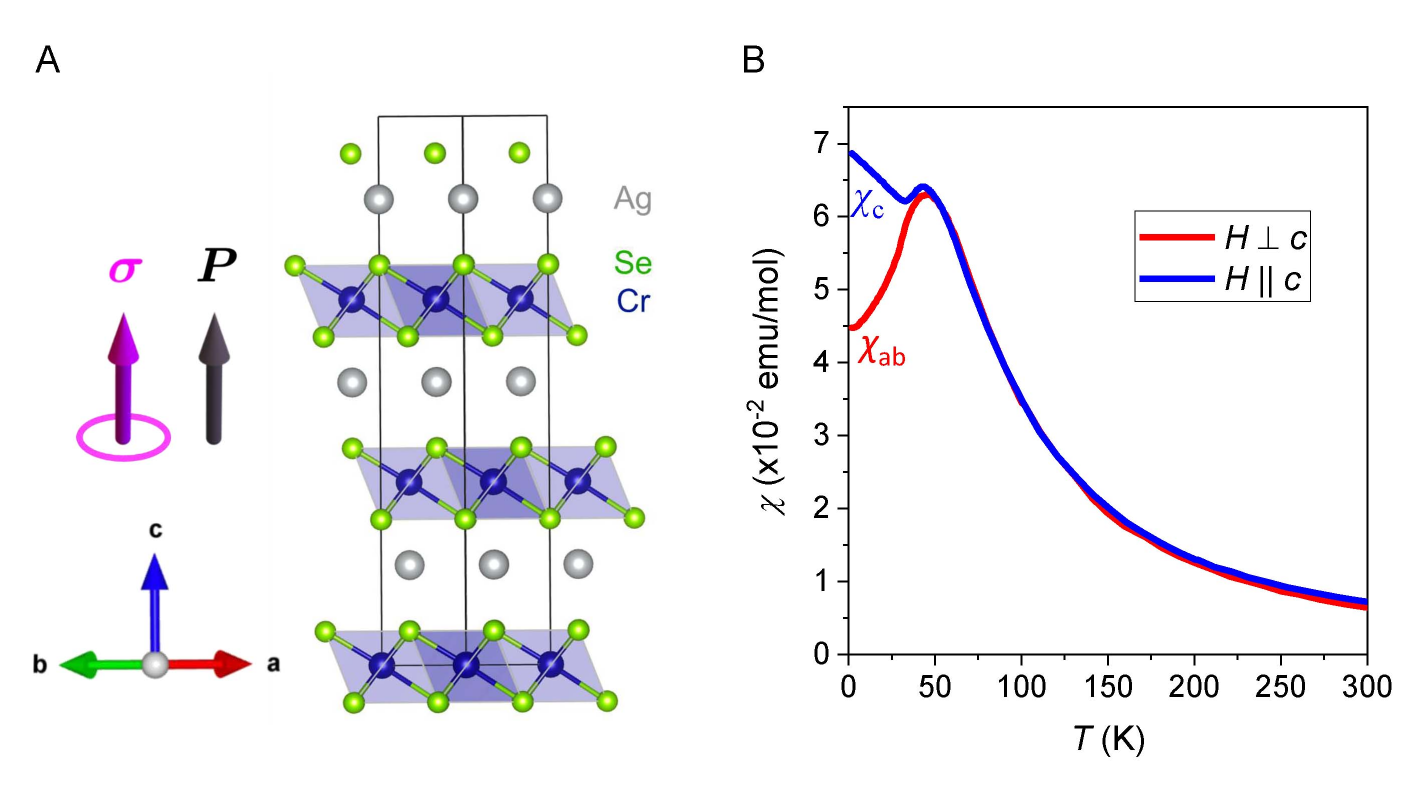}
\caption{Crystal structure and magnetic susceptibility of AgCrSe$_2$. (\emph{A}) Crystal structure of AgCrSe$_2$ (space group \emph{R3m}). The polarization direction $\boldsymbol{P}$ and the symmetry allowed Hall pseudovector $\boldsymbol{\sigma}$ (corresponding to an anomalous Hall conductivity $\sigma_{xy}$) are marked. (\emph{B}) Magnetic susceptibility measured with an applied 1 T magnetic field along and perpendicular to the \emph{c}-axis, respectively.
}
\label{fig1}
\end{figure}

\begin{figure*}[!t]
\centering
\includegraphics[width=0.9\linewidth]{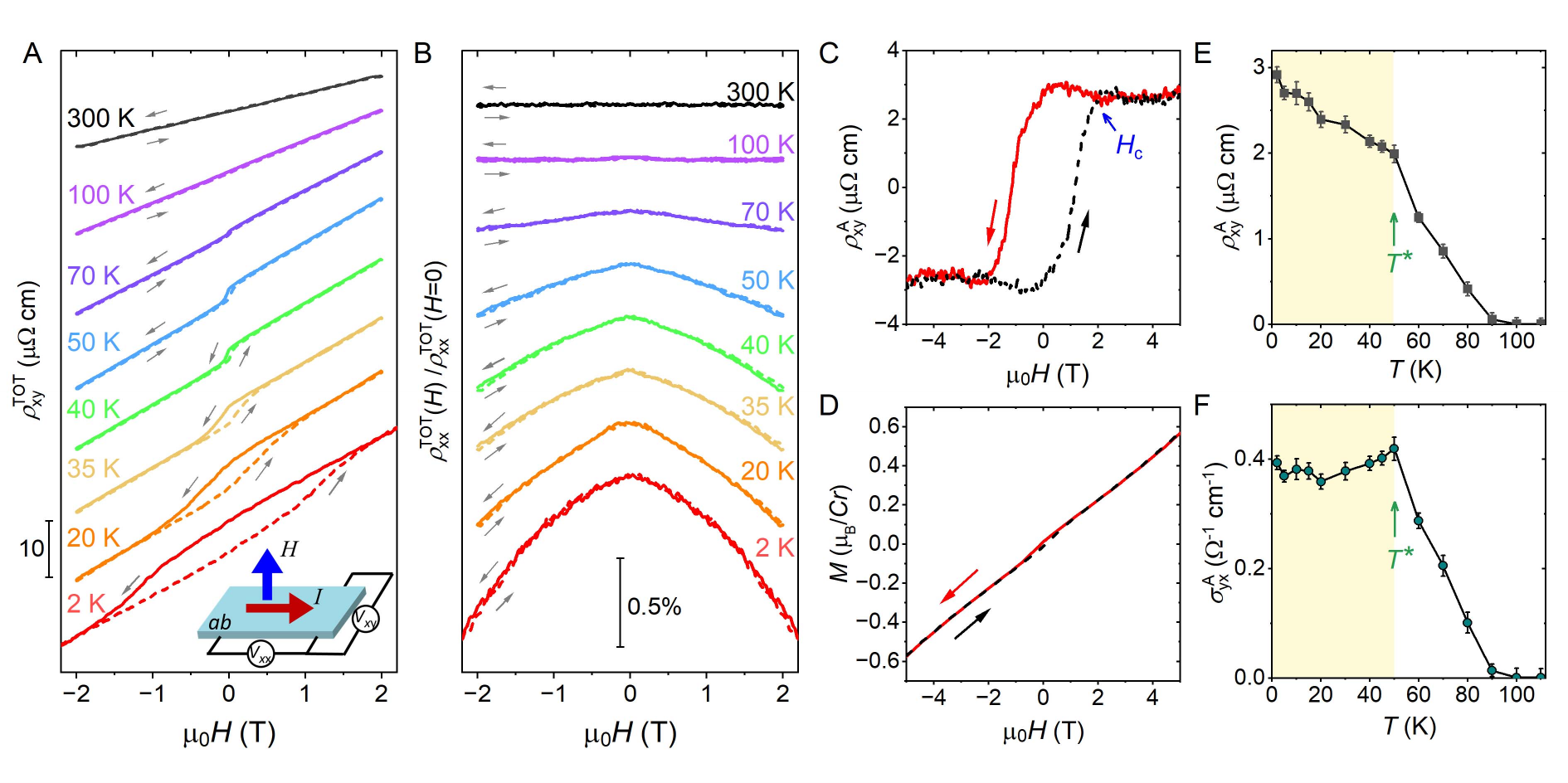}
\caption{Spontaneous AHE observed in AgCrSe$_2$. (\emph{A}) Hall resistivity ($\rho_{xy}^{TOT}(H)$), as well as (\emph{B}) normalized longitudinal resistivity ($\rho_{xx}^{TOT}(H)/\rho_{xx}^{TOT}(H=0)$), measured simultaneously with $H$ applied along the $c$-axis. The curves are offset vertically for clarity. Inset of (\emph{A}): a schematic illustrating the measurement setup. (\emph{C}) $\rho_{xy}^A= \rho_{xy}^{TOT} - \rho_{xy}^0$ is plotted as a function of $H$ at 2 K. The resistivity shows a clear jump and saturates at a critical field $H_c$. (\emph{D}) Magnetization $M$ vs $H$ measured at 2 K with the magnetic field applied along the $c$-axis. (\emph{E}) Temperature evolution of the zero-field component $\rho_{xy}^A (H=0)$ and (\emph{F}) the corresponding $\sigma_{yx}^A$ obtained by inverting the resistivity tensor. Error bars include the uncertainty in extracting the zero-field resistivity from the Hall measurements.  }
\label{fig2}
\end{figure*}

AgCrSe$_2$ has a layered structure with alternate Ag layers and edge-sharing CrSe\textsubscript{6} octahedral layers repeating along the $c$-axis~\cite{gascoin2011order,li2018liquid,PNASDing2020,Baenitz,PhysRevMaterials.6.054602}, as illustrated in Fig.~\ref{fig1}A. The compound crystallizes into the noncentrosymmetric \emph{R3m} space group. The polar structure is realized by the alternating layers of Ag and CrSe\textsubscript{6}, which break the inversion symmetry and allow for a polarization direction along the $c$-axis (Fig.~\ref{fig1}A)~\cite{PhysRevLett.101.067204, Singh2009, xiang2022}. The Cr atoms in each layer form a triangular lattice and host $S$ = 3/2 spins. Previous neutron diffraction characterization and magnetization measurements reveal that the adjacent octahedral layers couple antiferromagnetically. A noncollinear spin structure in the $ab$-plane has been reported  ~\cite{engelsman1973crystal,damay2016localised,Baenitz}, revealing interplay between various channels of intralayer exchange couplings. In Fig.~\ref{fig1}B we present the temperature dependence of the magnetic susceptibility $\chi$ of a AgCrSe$_2$ single crystal, measured with a magnetic field applied along the $c$-axis ($\chi_c$) and in the $ab$-plane ($\chi_{ab}$), respectively. The susceptibility is isotropic at high temperatures, but $\chi_{ab}$ and $\chi_c$ deviate from each other below $T^* = 50$ K, the characteristic temperature at which spin order begins to become long-ranged.

For the magneto-transport measurements, we realized microfabricated devices based on exfoliated AgCrSe$_2$ crystals with thicknesses ranging from 100 nm to 800 nm and compared our results with those from bulk single crystal devices. As expected in a material with a layered crystal structure, the resistive anisotropy $\rho_{zz}^{TOT}/\rho_{xx}^{TOT}$ is large, rising from 25 at 50 K to 100 at 2 K, and the in-plane resistivity at 2 K is approximately 2 $\textnormal {m} \Omega \, \textnormal {cm}$, which may be the result of doping due to an intrinsic non-stoichiometry. Throughout the paper we map the Cartesian coordinates commonly used to describe the Hall effect to the crystalline ones, with the $z$ axis being the crystalline $c$ axis, the $xy$ plane being the $ab$ plane of the crystalline layers. \\

We first discuss our main experimental evidence for the AHE, which was observed by measuring the Hall resistivity $\rho_{xy}^{TOT}$ upon applying the magnetic field $H$ perpendicular to the $ab$-plane. The measurements were carried out employing the setup as illustrated by the schematic in the inset of Fig.~2A. From 300 K to 100 K, the Hall resistivity $\rho_{xy}^{TOT}$ is linear as a function of $H$ (Fig.~\ref{fig2}A). The Hall coefficient is positive, indicating that the majority charge carriers in the system are holes. With further cooling, $\rho_{xy}^{TOT}$ exhibits a clear hysteresis loop with a sizable jump and width 2$H_c$, when the magnetic field is swept back and forth. The Hall resistivity jump and $H_c$ become more pronounced upon lowering the temperature. In contrast to the presence of large jumps in $\rho_{xy}^{TOT}$, the longitudinal resistivity $\rho_{xx}^{TOT}$ measured concomitantly does not exhibit significant variations in this temperature regime, as shown in Fig.~\ref{fig2}B.

\begin{figure}
\centering
\includegraphics[width=1\linewidth]{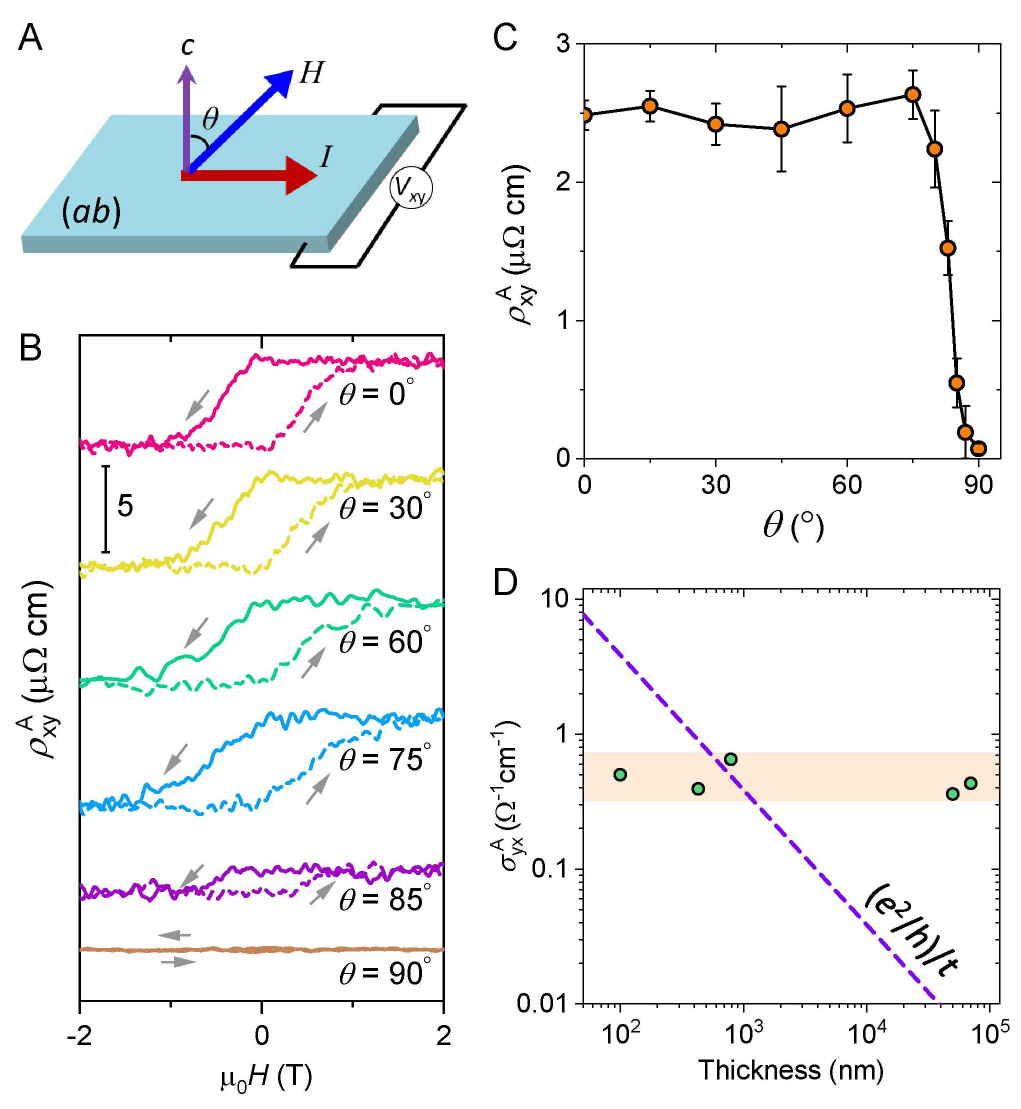}
\caption{Angular and thickness dependence of the AHE.(\emph{A}) Illustration of the measurement setup for the angular dependent AHE. The current is applied in the $ab$-plane, and the magnetic field is rotating relative to the $c$-axis. (\emph{B}) $\rho_{xy}^A(H)$ measured at different rotation angles at fixed temperature 5 K. (\emph{C}) Angular dependence of the zero-field resistivity $\rho_{xy}^A$. $\rho_{xy}^A$ remains a plateau up to $\theta \sim 80^\circ$, and then drops abruptly to near zero when $H$ is aligned to the $ab$-plane ($\theta=90^\circ$). Error bars reflect an estimate of the uncertainty in extracting the resistivity jumps. (\emph{D}) Thickness dependence of $\sigma_{yx}^A$. The dashed line is the quantum conductance $e^2/h$ normalized by the thickness $t$. 
}
\label{fig3}
\end{figure}

A common question regarding hysteretic AHE signals is their relationship to the magnetization.  Because the Hall effect can have a variety of origins, the total Hall conductivity $\sigma_{xy}^{TOT} $ can be expressed as a sum of contributions: $\sigma_{xy}^{TOT} =\sigma_{xy}^0 + \sigma_{xy}^A $, where $\sigma_{xy}^0$ is the traditional Hall conductivity from orbital electronic motion and $\sigma_{xy}^A$ is the term resulting from $k$-space Berry curvature.   Written in terms of the measured quantities which are resistivities, $\sigma_{yx}^{TOT}=\rho_{xy}^{TOT} / ((\rho_{xy}^{TOT})^2+ (\rho_{xx}^{TOT})^2 )$. If (and only if) $(\rho_{xx}^{TOT})^2 \gg  (\rho_{xy}^{TOT})^2 $ and $\rho_{xx}^{TOT}$ has a weak magnetic field dependence, a similar separation can be made, to a good approximation, in the Hall resistivity: 

\begin{equation} 
\rho_{xy}^{TOT} (H)= R_0 \mu_0 H + \rho_{xy}^A \,
\label{eq1}
\end{equation}

Here, $R_0$ is the ordinary Hall coefficient, $\mu_0$ is the permeability, and we use $ \rho_{xy}^0$ to represent $R_0 \mu_0 H$. In our AgCrSe$_2$ microcrystals, $\rho_{xx}^{TOT}$ varies between 1.5 and 3 $\textnormal {m} \Omega \, \textnormal {cm}$ between 2 K and 300 K, with a magnetoresistance of less than 2$\%$ for $\mu_0 H < 4$ T. Inspection of Fig.~\ref{fig2} shows that the above condition is therefore very well satisfied, so the decomposition of Eq.~\ref{eq1} is justified. 

The field-linear part of $\rho_{xy}^{TOT}(H)$ enables the identification of $R_0$, and the subtraction of $ \rho_{xy}^0$. As shown in Fig.~\ref{fig2}C,
the anomalous Hall resistivity, $ \rho_{xy}^A$, is seen to be hysteretic with $H_c$ = 2 T at 2 K and a sizable resistivity jump of $\cong 6 \mu\Omega$ cm.
We also measure the magnetization $M$ as a function of the applied field, and plot it in Fig.~\ref{fig2}D. Clearly the large hysteresis shown in $ \rho_{xy}^A$ (Fig.~\ref{fig2}C) cannot be explained by the linear contribution from $M$ (Fig.~\ref{fig2}D), which is distinct from the conventional ferromagnetism. \\

One diagnostic sometimes used for the existence of an intrinsic AHE is an anomalous Hall conductivity, $\sigma_{yx}^A$, that is independent of scattering~\cite{RevModPhys.82.1539}. 
In Fig.~\ref{fig2}E we show the temperature dependence of the zero-field component $\rho_{xy}^A (H=0)$. The corresponding $\sigma_{yx}^A (H=0)$ obtained by inverting the resistivity tensor $\sigma_{yx}^A = \rho_{xy}^A/((\rho_{xx}^{TOT})^2+(\rho_{xy}^A)^2)$, is plotted in Fig.~\ref{fig2}F, and $\sigma_{yx}^A$ is found to saturate. The fact that $\sigma_{yx}^A$ remains constant over a range of temperature in which $\sigma_{xx}^{TOT}$ and hence $\rho_{xy}^A$ are both temperature dependent is consistent with the scattering rate independence expected for a momentum-space Berry-curvature related effect. \\

\begin{figure*}[t]
\centering
\includegraphics[width=0.8\linewidth]{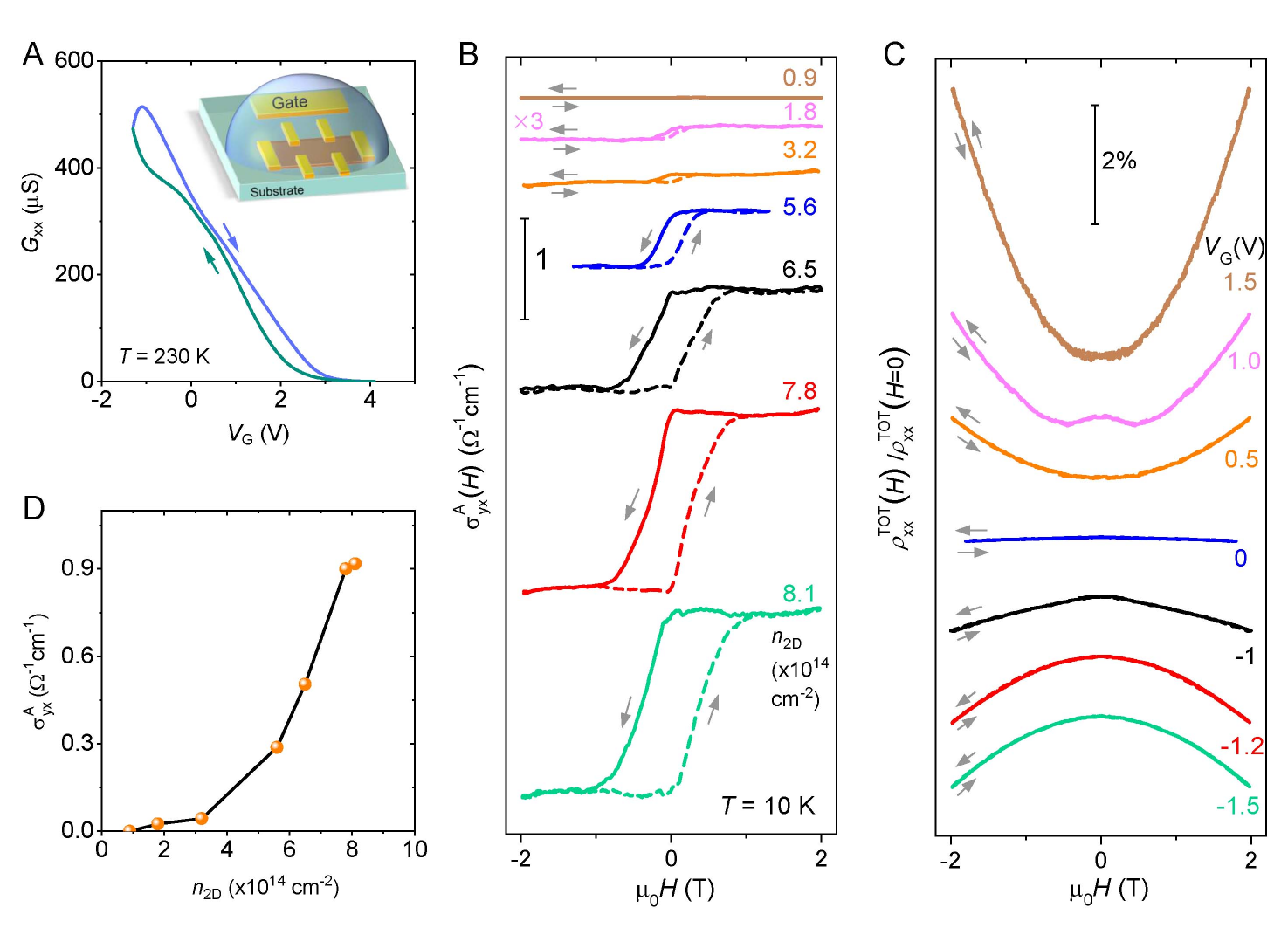}
\caption{Gate-tunable AHE in a AgCrSe$_2$ thin flake. 
(\emph{A}) Conductance as a function of the applied gate voltage measured at 230 K. The inset shows a schematic of the device measurement setup.  The ionic liquid (DEME-TFSI) covers both the side gate electrode and the thin flake. (\emph{B}) $\sigma_{yx}^A(H)$ and (\emph{C}) normalized longitudinal resistivity ($\rho_{xx}^{TOT}(H)/\rho_{xx}^{TOT}(H=0)$) modulated by the ionic gating. The data were obtained at $T$ = 10 K with the magnetic field applied along the $c$-axis. The applied $V_{\textnormal {G}}$ and the carrier density $n_{2D}$ extracted from the field linear part of the Hall effect at each $V_{G}$ are labeled beside the curves. (\emph{D}) $\sigma_{yx}^A $ as a function of the doping level.
}
\label{fig4}
\end{figure*}

The data shown in Fig.~\ref{fig2} establish one of our key experimental findings, namely the existence of an AHE in AgCrSe$_2$. In order to investigate further, we study the dependence of $\rho_{xy}^A$ on the angle of the magnetic field relative to the crystalline $c$-axis, rotating it by angle $\theta$ in the plane of the $c$-axis and the applied current, as shown in Fig.~\ref{fig3}A.  The result, shown in Fig.~\ref{fig3}B and summarized in Fig.~\ref{fig3}C, is striking: $\rho_{xy}^A$ remains approximately angle-independent before `switching off' for $\theta > 80^\circ$. At first sight, such an observation seems to be consistent with a topological response of some kind, in which $\rho_{xy}^A$ is fixed at a topologically-controlled value, insensitive to changes in external conditions.  However, the simplest interpretations of this kind are inconsistent with the size of the AHE that we observe. A topological contribution to the AHE in a two-dimensional system leads to a quantized conductance in units of $e^2/h$, where $e$ is the electronic charge and $h$ is the Planck constant.  Although our measured $\rho_{xy}^A$ is large---comparable with the largest seen in any antiferromagnetic system~\cite{nakatsuji2015large,nayak2016large,ghimire2018large,lee2019spin}---inverting the resistivity tensor and converting to two dimensions by factoring out the layer separation of 7 Å gives $\sigma_{yx}^A \cong 7 \times 10^{-4} e^2/h$ per atomic layer.  This low value of $\sigma_{yx}^A$ means that we are far from a topological response based on any bulk physics of AgCrSe$_2$.

However, one other form of topological response is worthy of consideration. A net non-zero Chern number due to surfaces could result in the reduction of the observed $\sigma_{yx}^A$ from the quantized value by the ratio $n_s/n_b$, where $n_s$ is the number of topologically non-trivial layers at or near the sample surface and $n_b$ is the number of topologically trivial bulk layers.  This intriguing hypothesis can be checked experimentally by studying samples of varying thickness.  We did this, with the results shown in Fig.~\ref{fig3}D.  The observed $\sigma_{yx}^A$ shows no systematic dependence on sample thickness, and hence none on $n_s/n_b$.  We therefore believe that our observation is that of a non-quantized $\sigma_{yx}^A$ whose origin is in the bulk physics of AgCrSe$_2$.

\begin{figure*}
\centering
\includegraphics[width=0.9\linewidth]{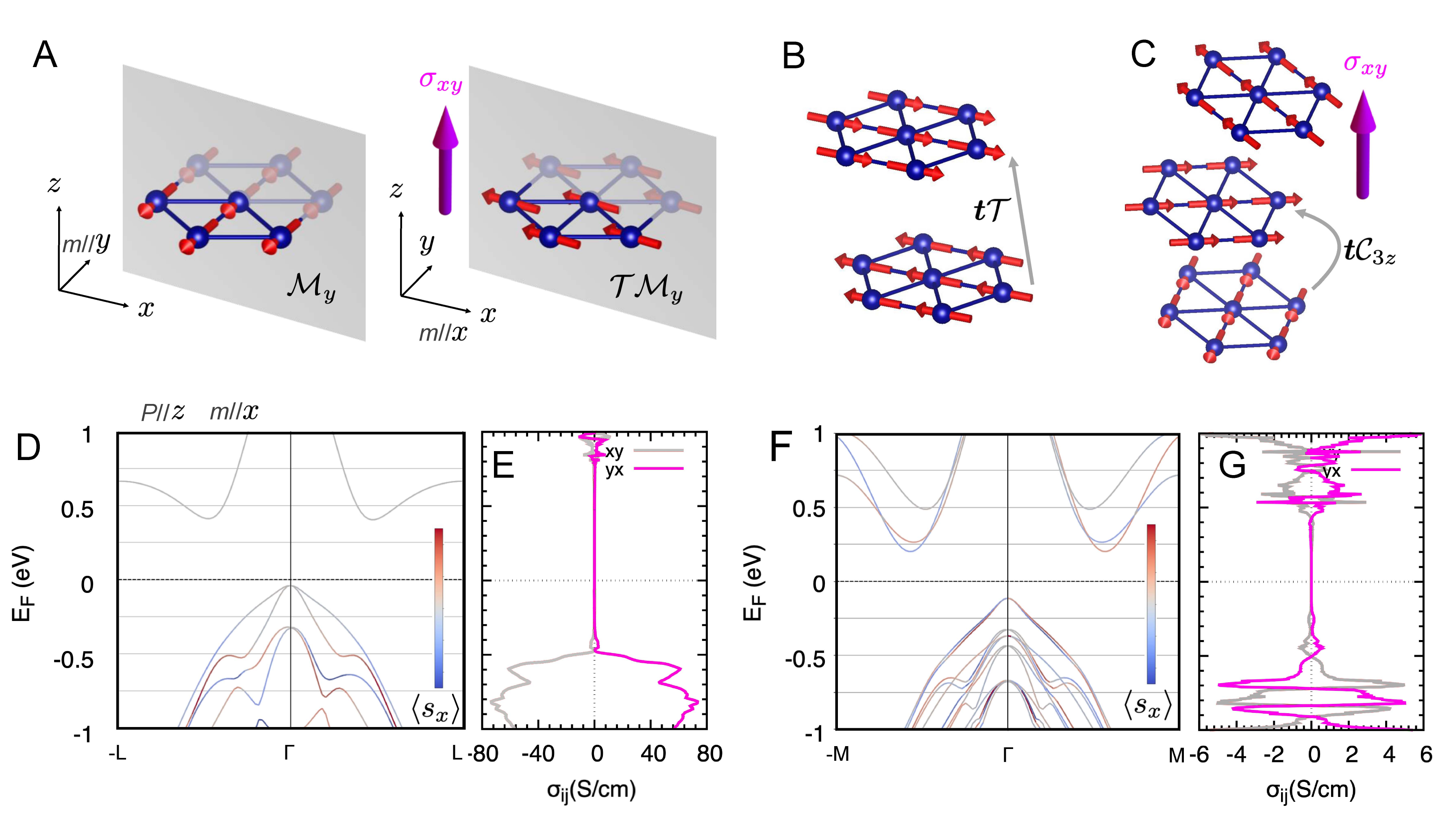}
\caption{Theoretical calculations of AHE in AgCrSe$_2$. 
(\emph{A}) Model of ferromagnetic Cr atoms in a monolayer. The mirror symmetry plane $\mathcal{M}_y$ is marked in gray color. The mirror symmetry translation coupled with time reversal symmetry $\mathcal{T}\mathcal{M}_y$ allows for a Hall vector (right panel). (\emph{B}) Model of Cr atoms in the collinear antiferromagnetic states. The unit cell translation coupled with time reversal $t\mathcal{T}$ cancels the Hall vector. (\emph{C}) Model of Cr atoms in the noncollinear states. The unit cell translation coupled with rotation symmetry $t\mathcal{C}_{3z}$ is marked. (\emph{D}, \emph{E})) Spin-projected energy bands (\emph{D}) and energy dependent anomalous Hall conductivity (\emph{E}) in ferromagnetic states with spins aligned parallel to the $x$-axis. (\emph{F}, \emph{G})) Spin-projected energy bands (\emph{F}) and energy dependent anomalous Hall conductivity (\emph{G}) in antiferromagnetic states in a simplified noncollinear structure. The calculations, for stoichimetric AgCrSe$_2$, put the Fermi level in the band gap. In the real crystals, non-stoichiometry places it in the hole bands, where the calculations give a finite $\sigma_{xy}$ with a strong doping dependence, qualitatively in accord with the results of our gating experiments.
}
\label{fig5}
\end{figure*}

Another, even more direct, way to investigate the surface layers of a material such as AgCrSe$_2$ is to apply an ionic gate using an ionic liquid. We now demonstrate that an ionic gate  drastically modulates the AHE. We employ an ionic field effect transistor setup, in the configuration schematically illustrated in the inset of Fig.~\ref{fig4}A. The device includes an exfoliated AgCrSe$_2$ thin flake, a large-area side gate pad, as well as the ionic liquid that covers both the thin flake and the gate electrode. Due to the screening effect, the induced conductivity takes place in the surface layers of the material. Fig.~\ref{fig4}A shows the in-plane conductance tuned as a function of applied gate voltage, $V_G$. A moderate $V_G$ of a few volts can change the conductance by orders of magnitude. The high doping level of an ionic gate also leads to a drastic change in the AHE, and we show the results of such experiments in Figs.~\ref{fig4}B-D. 

In the experiment, we applied both negative and positive gate voltages, producing sheet carrier densities per layer that are both smaller and larger than the carrier density per layer of the bulk devices. As is seen in Fig.~\ref{fig4}B, an AHE with similar characteristics to that of the bulk is observed, with a magnitude that is tuned to be larger (at densities of 6.5, 7.8 and 8.1 $\times$ 10$^{14}$ cm$^{-2}$) and smaller than that observed in the bulk.  As with the bulk case, the pronounced hysteresis seen in the AHE is absent in the magnetoresistance, shown in Fig.~\ref{fig4}C. The AHE is modulated to be vanishing at carrier densities smaller than 1.8 $\times$ 10$^{14}$ cm$^{-2}$.\\

By isolating the carriers to a quasi-two-dimensional sheet at the surface of the AgCrSe$_2$ crystals, and demonstrating that $\sigma_{yx}^A $ remains far from a quantized value, we provide further strong evidence for a non-quantized anomalous Hall origin of the signal. We will show below, through symmetry arguments and density functional theory (DFT) calculations, that its origin is still Berry curvature, but a carrier-density-dependent one in the valence band of these $p$-type doped crystals.

We start from the symmetry analysis to look for necessary conditions for the occurrence of a nonzero anomalous Hall conductivity, i.e. $\sigma_{xy}$. The out-of-plane spin tilting that is favored by Dzyaloshinsky-Moriya interaction~\cite{DZYALOSHINSKY1958241, PhysRev.120.91} can, in principle, break symmetries in the crystal and give rise to a ferromagnetic behavior. However, as previously discussed, the resulting net moment alone cannot fully explain the observed large hysteresis loop in the Hall resistivity, and an additional component in the Hall resistivity is still observed after taking the moment into account. Therefore, we resort to other necessary conditions that can possibly explain the observed phenomena. 

Inspection of the magnetic space group for various possible magnetic orderings reveals that an anomalous Hall conductivity can be generated either by (a) an in-plane magnetic moment ~\cite{PhysRevLett.111.086802} and/or (b) by the noncollinear spin structure.  As shown in the left panel of Fig.~\ref{fig5}A, for spins parallel to the $y$-axis ($m \parallelsum y$), the mirror symmetry $\mathcal{M}_y$ is retained which excludes the Hall vector. It is worth noting that the Hall conductivity components  correspond to the antisymmetric part of the 2nd rank conductivity tensor and thus we can write them as components of a pseudovector, Hall vector, $\boldsymbol{\sigma}= (\sigma_{yz},\sigma_{zx},\sigma_{xy})$~\cite{vsmejkal2022anomalous,landau1995electrodynamics}. In contrast, for spins parallel to the $x$-axis ($m \parallelsum x$), the mirror symmetry is augmented by time-reversal $\mathcal{T}\mathcal{M}_y$ which allows for a Hall vector component perpendicular to the $xy$-plane (right panel of Fig.~\ref{fig5}A).  Furthermore, we consider two types of antiferromagnetic states and show that while simple collinear antiferromagnetic ordering cannot explain the observed signal, a nonzero anomalous Hall conductivity is anticipated in the noncollinear antiferromagnetic states for the AgCrSe$_2$ crystal structure. We construct the collinear antiferromagnetic state by doubling of the unit cell along the $c$-axis. In the collinear antiferromagnetic states, the Cr sublattices are connected by the unit cell translation combined with time-reversal, $t \mathcal{T}$ symmetry, as shown in Fig.~\ref{fig5}B, which forces the Hall vector to vanish. 
In contrast, the noncollinear antiferromagnetic state (inherent to the experimentally indicated cycloidal spin structure~\cite{Baenitz}]) can break TRS in electronic structure and lift Kramers spin degeneracy~\cite{PhysRevLett.112.017205}. 
To emulate the effect of noncollinear antiferromagnetic ordering we have considered in our calculations a simplified noncollinear spin structure, as shown in Fig.~\ref{fig5}C, constructed by tripling the unit cell along the $c$-axis. The noncollinear antiferromagnetic sublattices are, in this case, related by three fold rotation combined with translation, $t\mathcal{C}_{3z}$, and thus 
allow a Hall vector perpendicular to the $xy$-plane~\cite{vsmejkal2020crystal,vsmejkal2022anomalous}. 
\\

To further investigate the interplay between the magnetic order and the polar structure, as well as to corroborate the aforementioned symmetry analysis, we performed DFT calculations including SOC and calculated the anomalous Hall conductivity based on the relativistic DFT band structures.
The obtained results for various spin configurations are plotted in Figs.~\ref{fig5}D-G. For spins aligned parallel to the $x$-axis (right panel of Fig.~\ref{fig5}A), the energy bands, as shown in Fig.~\ref{fig5}D, are spin split along the high-symmetry axes and asymmetric to $\Gamma$ point. This asymmetric band dispersion is a direct manifestation of the interplay between the polarization and the magnetic order: the polar structure introduces a Rashba-like spin-splitting and the magnetic moment along the $x$-axis ($m \parallelsum x$) leads to an asymmetric deformation. As a result of the band dispersion asymmetry, TRS is broken and an anomalous Hall conductivity up to 80 S/cm is obtained for energy levels lower in the valence band (Fig.~\ref{fig5}E). In addition, for the simplified noncollinear spin structure in the antiferromagnetic states, as shown in Fig.~\ref{fig5}C, the bands once again exhibit TRS breaking. The calculations yield the order of magnitude of $\sigma_{xy}$ consistent with the experimental data and capture Hall conductivity increase with a lowering Fermi level towards the bottom of the valence bands in the range probed by the gating experiments.

The above-described models are based on coplanar spin configurations which generate a Hall vector perpendicular to the plane. This is distinct from the conventional ferromagnetism that a Hall vector is generated by a moment along the same direction. The scenario involving a more complex noncollinear spin structure has been examined on model level, and is consistent with the main conclusions based on DFT results. We emphasize that these models discuss possible origins of the Hall signal only. 
Nevertheless, our theoretical analysis and calculations demonstrate that the AgCrSe$_2$ crystal structure family can host intriguing in-plane magnetization~\cite{PhysRevLett.111.086802} or noncollinear magnetism~\cite{PhysRevLett.112.017205} driven AHE.  
It also demonstrates that the kind of bands that exist in AgCrSe$_2$ can yield the anomalous Hall signals with a magnitude consistent with that we have observed, and thus the layered crystal structure allows for the coexistence of polar and TRS broken effects.\\

In conclusion, we have observed a spontaneous AHE in the polar, layered, triangular-lattice material AgCrSe$_2$, and have shown how its magnitude can be tuned by an ionic gate.  Although the measured anomalous Hall resistivity is comparable with the largest observed in any magnetic material, the anomalous Hall conductivity is far from the quantized value.  We show that it can be substantiated by symmetry analysis and DFT calculations.  However, the microscopic origin of the TRS breaking remains to be understood, as does the hysteresis in the AHE.  Our work motivates further theoretical investigation and detailed neutron scattering studies of this fascinating material class, which we have demonstrated to be an appealing candidate for the coexistence of ferroelectric- and ferromagnetic-like responses and functionalities. 
Moreover, the gate-controllable AHE in this material provides a new possibility for the local manipulation of spin states, which can facilitate the realization of stable and compact spintronic and magnetoelectric devices.\\

\section*{Acknowledgments}
We thank B. Doucot, P.D.C. King, B. Schmidt and V. Sunko for useful discussions, and S. Seifert for experimental support. Research in Dresden benefits from the environment provided by the DFG Cluster of Excellence ct.qmat (EXC 2147, project ID 390858940). S.-J.K. acknowledges support from the International Max Planck Research School for Chemistry and Physics of Quantum Materials (IMPRS-CPQM). LŠ acknowledges support from Johannes Gutenberg University Grant TopDyn, and support by the Deutsche Forschungsgemein- schaft (DFG, German Research Foundation) for funding through TRR 288 – 422213477 (projects A09 and B05). 
We thank U. Nitzsche for technical support. We are grateful to the Max Planck Society for financial support.

\bibliography{AHE_reference.bib}

\begin{thebibliography}{37}%
\makeatletter
\providecommand \@ifxundefined [1]{%
 \@ifx{#1\undefined}
}%
\providecommand \@ifnum [1]{%
 \ifnum #1\expandafter \@firstoftwo
 \else \expandafter \@secondoftwo
 \fi
}%
\providecommand \@ifx [1]{%
 \ifx #1\expandafter \@firstoftwo
 \else \expandafter \@secondoftwo
 \fi
}%
\providecommand \natexlab [1]{#1}%
\providecommand \enquote  [1]{``#1''}%
\providecommand \bibnamefont  [1]{#1}%
\providecommand \bibfnamefont [1]{#1}%
\providecommand \citenamefont [1]{#1}%
\providecommand \href@noop [0]{\@secondoftwo}%
\providecommand \href [0]{\begingroup \@sanitize@url \@href}%
\providecommand \@href[1]{\@@startlink{#1}\@@href}%
\providecommand \@@href[1]{\endgroup#1\@@endlink}%
\providecommand \@sanitize@url [0]{\catcode `\\12\catcode `\$12\catcode
  `\&12\catcode `\#12\catcode `\^12\catcode `\_12\catcode `\%12\relax}%
\providecommand \@@startlink[1]{}%
\providecommand \@@endlink[0]{}%
\providecommand \url  [0]{\begingroup\@sanitize@url \@url }%
\providecommand \@url [1]{\endgroup\@href {#1}{\urlprefix }}%
\providecommand \urlprefix  [0]{URL }%
\providecommand \Eprint [0]{\href }%
\providecommand \doibase [0]{https://doi.org/}%
\providecommand \selectlanguage [0]{\@gobble}%
\providecommand \bibinfo  [0]{\@secondoftwo}%
\providecommand \bibfield  [0]{\@secondoftwo}%
\providecommand \translation [1]{[#1]}%
\providecommand \BibitemOpen [0]{}%
\providecommand \bibitemStop [0]{}%
\providecommand \bibitemNoStop [0]{.\EOS\space}%
\providecommand \EOS [0]{\spacefactor3000\relax}%
\providecommand \BibitemShut  [1]{\csname bibitem#1\endcsname}%
\let\auto@bib@innerbib\@empty
\bibitem [{\citenamefont {Nagaosa}\ \emph {et~al.}(2010)\citenamefont
  {Nagaosa}, \citenamefont {Sinova}, \citenamefont {Onoda}, \citenamefont
  {MacDonald},\ and\ \citenamefont {Ong}}]{RevModPhys.82.1539}%
  \BibitemOpen
  \bibfield  {author} {\bibinfo {author} {\bibfnamefont {N.}~\bibnamefont
  {Nagaosa}}, \bibinfo {author} {\bibfnamefont {J.}~\bibnamefont {Sinova}},
  \bibinfo {author} {\bibfnamefont {S.}~\bibnamefont {Onoda}}, \bibinfo
  {author} {\bibfnamefont {A.~H.}\ \bibnamefont {MacDonald}},\ and\ \bibinfo
  {author} {\bibfnamefont {N.~P.}\ \bibnamefont {Ong}},\ }\bibfield  {title}
  {\bibinfo {title} {Anomalous {H}all effect},\ }\href
  {https://doi.org/10.1103/RevModPhys.82.1539} {\bibfield  {journal} {\bibinfo
  {journal} {Rev. Mod. Phys.}\ }\textbf {\bibinfo {volume} {82}},\ \bibinfo
  {pages} {1539} (\bibinfo {year} {2010})}\BibitemShut {NoStop}%
\bibitem [{\citenamefont {{\v{S}}mejkal}\ \emph {et~al.}(2022)\citenamefont
  {{\v{S}}mejkal}, \citenamefont {MacDonald}, \citenamefont {Sinova},
  \citenamefont {Nakatsuji},\ and\ \citenamefont
  {Jungwirth}}]{vsmejkal2022anomalous}%
  \BibitemOpen
  \bibfield  {author} {\bibinfo {author} {\bibfnamefont {L.}~\bibnamefont
  {{\v{S}}mejkal}}, \bibinfo {author} {\bibfnamefont {A.~H.}\ \bibnamefont
  {MacDonald}}, \bibinfo {author} {\bibfnamefont {J.}~\bibnamefont {Sinova}},
  \bibinfo {author} {\bibfnamefont {S.}~\bibnamefont {Nakatsuji}},\ and\
  \bibinfo {author} {\bibfnamefont {T.}~\bibnamefont {Jungwirth}},\ }\bibfield
  {title} {\bibinfo {title} {Anomalous {H}all antiferromagnets},\ }\href@noop
  {} {\bibfield  {journal} {\bibinfo  {journal} {Nat. Rev. Mater.}\ }\textbf
  {\bibinfo {volume} {7}},\ \bibinfo {pages} {482–496} (\bibinfo {year}
  {2022})}\BibitemShut {NoStop}%
\bibitem [{\citenamefont {Xiao}\ \emph {et~al.}(2010)\citenamefont {Xiao},
  \citenamefont {Chang},\ and\ \citenamefont {Niu}}]{RevModPhys.82.1959}%
  \BibitemOpen
  \bibfield  {author} {\bibinfo {author} {\bibfnamefont {D.}~\bibnamefont
  {Xiao}}, \bibinfo {author} {\bibfnamefont {M.-C.}\ \bibnamefont {Chang}},\
  and\ \bibinfo {author} {\bibfnamefont {Q.}~\bibnamefont {Niu}},\ }\bibfield
  {title} {\bibinfo {title} {Berry phase effects on electronic properties},\
  }\href {https://doi.org/10.1103/RevModPhys.82.1959} {\bibfield  {journal}
  {\bibinfo  {journal} {Rev. Mod. Phys.}\ }\textbf {\bibinfo {volume} {82}},\
  \bibinfo {pages} {1959} (\bibinfo {year} {2010})}\BibitemShut {NoStop}%
\bibitem [{\citenamefont {{\v{S}}mejkal}\ \emph {et~al.}(2020)\citenamefont
  {{\v{S}}mejkal}, \citenamefont {Gonz{\'a}lez-Hern{\'a}ndez}, \citenamefont
  {Jungwirth},\ and\ \citenamefont {Sinova}}]{vsmejkal2020crystal}%
  \BibitemOpen
  \bibfield  {author} {\bibinfo {author} {\bibfnamefont {L.}~\bibnamefont
  {{\v{S}}mejkal}}, \bibinfo {author} {\bibfnamefont {R.}~\bibnamefont
  {Gonz{\'a}lez-Hern{\'a}ndez}}, \bibinfo {author} {\bibfnamefont
  {T.}~\bibnamefont {Jungwirth}},\ and\ \bibinfo {author} {\bibfnamefont
  {J.}~\bibnamefont {Sinova}},\ }\bibfield  {title} {\bibinfo {title} {Crystal
  time-reversal symmetry breaking and spontaneous {H}all effect in collinear
  antiferromagnets},\ }\href@noop {} {\bibfield  {journal} {\bibinfo  {journal}
  {Sci. Adv.}\ }\textbf {\bibinfo {volume} {6}},\ \bibinfo {pages} {eaaz8809}
  (\bibinfo {year} {2020})}\BibitemShut {NoStop}%
\bibitem [{\citenamefont {Samanta}\ \emph {et~al.}(2020)\citenamefont
  {Samanta}, \citenamefont {Ležaić}, \citenamefont {Merte}, \citenamefont
  {Freimuth}, \citenamefont {Blügel},\ and\ \citenamefont
  {Mokrousov}}]{10.1063/5.0005017}%
  \BibitemOpen
  \bibfield  {author} {\bibinfo {author} {\bibfnamefont {K.}~\bibnamefont
  {Samanta}}, \bibinfo {author} {\bibfnamefont {M.}~\bibnamefont {Ležaić}},
  \bibinfo {author} {\bibfnamefont {M.}~\bibnamefont {Merte}}, \bibinfo
  {author} {\bibfnamefont {F.}~\bibnamefont {Freimuth}}, \bibinfo {author}
  {\bibfnamefont {S.}~\bibnamefont {Blügel}},\ and\ \bibinfo {author}
  {\bibfnamefont {Y.}~\bibnamefont {Mokrousov}},\ }\bibfield  {title} {\bibinfo
  {title} {{Crystal {H}all and crystal magneto-optical effect in thin films of
  {S}r{R}u{O}$_3$}},\ }\href@noop {} {\bibfield  {journal} {\bibinfo  {journal}
  {J. Appl. Phys.}\ }\textbf {\bibinfo {volume} {127}} (\bibinfo {year}
  {2020})}\BibitemShut {NoStop}%
\bibitem [{\citenamefont {Mazin}\ \emph {et~al.}(2021)\citenamefont {Mazin},
  \citenamefont {Koepernik}, \citenamefont {Johannes}, \citenamefont
  {González-Hernández},\ and\ \citenamefont {Šmejkal}}]{Mazin2021}%
  \BibitemOpen
  \bibfield  {author} {\bibinfo {author} {\bibfnamefont {I.~I.}\ \bibnamefont
  {Mazin}}, \bibinfo {author} {\bibfnamefont {K.}~\bibnamefont {Koepernik}},
  \bibinfo {author} {\bibfnamefont {M.~D.}\ \bibnamefont {Johannes}}, \bibinfo
  {author} {\bibfnamefont {R.}~\bibnamefont {González-Hernández}},\ and\
  \bibinfo {author} {\bibfnamefont {L.}~\bibnamefont {Šmejkal}},\ }\bibfield
  {title} {\bibinfo {title} {Prediction of unconventional magnetism in doped
  {F}e{S}b$_2$},\ }\href {https://doi.org/10.1073/pnas.2108924118} {\bibfield
  {journal} {\bibinfo  {journal} {Proc. Natl. Acad. Sci. U.S.A.}\ }\textbf
  {\bibinfo {volume} {118}},\ \bibinfo {pages} {e2108924118} (\bibinfo {year}
  {2021})}\BibitemShut {NoStop}%
\bibitem [{\citenamefont {Guin}\ \emph {et~al.}(2021)\citenamefont {Guin},
  \citenamefont {Xu}, \citenamefont {Kumar}, \citenamefont {Kung},
  \citenamefont {Dufresne}, \citenamefont {Le}, \citenamefont {Vir},
  \citenamefont {Michiardi}, \citenamefont {Pedersen}, \citenamefont
  {Gorovikov} \emph {et~al.}}]{guin20212d}%
  \BibitemOpen
  \bibfield  {author} {\bibinfo {author} {\bibfnamefont {S.~N.}\ \bibnamefont
  {Guin}}, \bibinfo {author} {\bibfnamefont {Q.}~\bibnamefont {Xu}}, \bibinfo
  {author} {\bibfnamefont {N.}~\bibnamefont {Kumar}}, \bibinfo {author}
  {\bibfnamefont {H.-H.}\ \bibnamefont {Kung}}, \bibinfo {author}
  {\bibfnamefont {S.}~\bibnamefont {Dufresne}}, \bibinfo {author}
  {\bibfnamefont {C.}~\bibnamefont {Le}}, \bibinfo {author} {\bibfnamefont
  {P.}~\bibnamefont {Vir}}, \bibinfo {author} {\bibfnamefont {M.}~\bibnamefont
  {Michiardi}}, \bibinfo {author} {\bibfnamefont {T.}~\bibnamefont {Pedersen}},
  \bibinfo {author} {\bibfnamefont {S.}~\bibnamefont {Gorovikov}}, \emph
  {et~al.},\ }\bibfield  {title} {\bibinfo {title}
  {2{D}-{B}erry-curvature-driven large anomalous {H}all effect in layered
  topological nodal-line mnalge},\ }\href@noop {} {\bibfield  {journal}
  {\bibinfo  {journal} {Adv. Mater.}\ }\textbf {\bibinfo {volume} {33}},\
  \bibinfo {pages} {2006301} (\bibinfo {year} {2021})}\BibitemShut {NoStop}%
\bibitem [{\citenamefont {\ifmmode~\check{S}\else \v{S}\fi{}mejkal}\ \emph
  {et~al.}(2022{\natexlab{a}})\citenamefont {\ifmmode~\check{S}\else
  \v{S}\fi{}mejkal}, \citenamefont {Sinova},\ and\ \citenamefont
  {Jungwirth}}]{PhysRevX.12.040501}%
  \BibitemOpen
  \bibfield  {author} {\bibinfo {author} {\bibfnamefont {L.}~\bibnamefont
  {\ifmmode~\check{S}\else \v{S}\fi{}mejkal}}, \bibinfo {author} {\bibfnamefont
  {J.}~\bibnamefont {Sinova}},\ and\ \bibinfo {author} {\bibfnamefont
  {T.}~\bibnamefont {Jungwirth}},\ }\bibfield  {title} {\bibinfo {title}
  {Emerging research landscape of altermagnetism},\ }\href
  {https://doi.org/10.1103/PhysRevX.12.040501} {\bibfield  {journal} {\bibinfo
  {journal} {Phys. Rev. X}\ }\textbf {\bibinfo {volume} {12}},\ \bibinfo
  {pages} {040501} (\bibinfo {year} {2022}{\natexlab{a}})}\BibitemShut
  {NoStop}%
\bibitem [{\citenamefont {\ifmmode~\check{S}\else \v{S}\fi{}mejkal}\ \emph
  {et~al.}(2022{\natexlab{b}})\citenamefont {\ifmmode~\check{S}\else
  \v{S}\fi{}mejkal}, \citenamefont {Sinova},\ and\ \citenamefont
  {Jungwirth}}]{PhysRevX.12.031042}%
  \BibitemOpen
  \bibfield  {author} {\bibinfo {author} {\bibfnamefont {L.}~\bibnamefont
  {\ifmmode~\check{S}\else \v{S}\fi{}mejkal}}, \bibinfo {author} {\bibfnamefont
  {J.}~\bibnamefont {Sinova}},\ and\ \bibinfo {author} {\bibfnamefont
  {T.}~\bibnamefont {Jungwirth}},\ }\bibfield  {title} {\bibinfo {title}
  {Beyond conventional ferromagnetism and antiferromagnetism: A phase with
  nonrelativistic spin and crystal rotation symmetry},\ }\href
  {https://doi.org/10.1103/PhysRevX.12.031042} {\bibfield  {journal} {\bibinfo
  {journal} {Phys. Rev. X}\ }\textbf {\bibinfo {volume} {12}},\ \bibinfo
  {pages} {031042} (\bibinfo {year} {2022}{\natexlab{b}})}\BibitemShut
  {NoStop}%
\bibitem [{\citenamefont {Gonzalez~Betancourt}\ \emph
  {et~al.}(2023)\citenamefont {Gonzalez~Betancourt}, \citenamefont
  {Zub\'a\ifmmode~\check{c}\else \v{c}\fi{}}, \citenamefont
  {Gonzalez-Hernandez}, \citenamefont {Geishendorf}, \citenamefont {\ifmmode
  \check{S}\else \v{S}\fi{}ob\'a\ifmmode~\check{n}\else \v{n}\fi{}},
  \citenamefont {Springholz}, \citenamefont {Olejn\'{\i}k}, \citenamefont
  {\ifmmode~\check{S}\else \v{S}\fi{}mejkal}, \citenamefont {Sinova},
  \citenamefont {Jungwirth}, \citenamefont {Goennenwein}, \citenamefont
  {Thomas}, \citenamefont {Reichlov\'a}, \citenamefont {\ifmmode~\check{Z}\else
  \v{Z}\fi{}elezn\'y},\ and\ \citenamefont
  {Kriegner}}]{PhysRevLett.130.036702}%
  \BibitemOpen
  \bibfield  {author} {\bibinfo {author} {\bibfnamefont {R.~D.}\ \bibnamefont
  {Gonzalez~Betancourt}}, \bibinfo {author} {\bibfnamefont {J.}~\bibnamefont
  {Zub\'a\ifmmode~\check{c}\else \v{c}\fi{}}}, \bibinfo {author} {\bibfnamefont
  {R.}~\bibnamefont {Gonzalez-Hernandez}}, \bibinfo {author} {\bibfnamefont
  {K.}~\bibnamefont {Geishendorf}}, \bibinfo {author} {\bibfnamefont
  {Z.}~\bibnamefont {\ifmmode \check{S}\else
  \v{S}\fi{}ob\'a\ifmmode~\check{n}\else \v{n}\fi{}}}, \bibinfo {author}
  {\bibfnamefont {G.}~\bibnamefont {Springholz}}, \bibinfo {author}
  {\bibfnamefont {K.}~\bibnamefont {Olejn\'{\i}k}}, \bibinfo {author}
  {\bibfnamefont {L.}~\bibnamefont {\ifmmode~\check{S}\else \v{S}\fi{}mejkal}},
  \bibinfo {author} {\bibfnamefont {J.}~\bibnamefont {Sinova}}, \bibinfo
  {author} {\bibfnamefont {T.}~\bibnamefont {Jungwirth}}, \bibinfo {author}
  {\bibfnamefont {S.~T.~B.}\ \bibnamefont {Goennenwein}}, \bibinfo {author}
  {\bibfnamefont {A.}~\bibnamefont {Thomas}}, \bibinfo {author} {\bibfnamefont
  {H.}~\bibnamefont {Reichlov\'a}}, \bibinfo {author} {\bibfnamefont
  {J.}~\bibnamefont {\ifmmode~\check{Z}\else \v{Z}\fi{}elezn\'y}},\ and\
  \bibinfo {author} {\bibfnamefont {D.}~\bibnamefont {Kriegner}},\ }\bibfield
  {title} {\bibinfo {title} {Spontaneous anomalous {H}all effect arising from
  an unconventional compensated magnetic phase in a semiconductor},\ }\href
  {https://doi.org/10.1103/PhysRevLett.130.036702} {\bibfield  {journal}
  {\bibinfo  {journal} {Phys. Rev. Lett.}\ }\textbf {\bibinfo {volume} {130}},\
  \bibinfo {pages} {036702} (\bibinfo {year} {2023})}\BibitemShut {NoStop}%
\bibitem [{\citenamefont {Feng}\ \emph {et~al.}(2022)\citenamefont {Feng},
  \citenamefont {Zhou}, \citenamefont {{\v{S}}mejkal}, \citenamefont {Wu},
  \citenamefont {Zhu}, \citenamefont {Guo}, \citenamefont
  {Gonz{\'a}lez-Hern{\'a}ndez}, \citenamefont {Wang}, \citenamefont {Yan},
  \citenamefont {Qin} \emph {et~al.}}]{feng2022anomalous}%
  \BibitemOpen
  \bibfield  {author} {\bibinfo {author} {\bibfnamefont {Z.}~\bibnamefont
  {Feng}}, \bibinfo {author} {\bibfnamefont {X.}~\bibnamefont {Zhou}}, \bibinfo
  {author} {\bibfnamefont {L.}~\bibnamefont {{\v{S}}mejkal}}, \bibinfo {author}
  {\bibfnamefont {L.}~\bibnamefont {Wu}}, \bibinfo {author} {\bibfnamefont
  {Z.}~\bibnamefont {Zhu}}, \bibinfo {author} {\bibfnamefont {H.}~\bibnamefont
  {Guo}}, \bibinfo {author} {\bibfnamefont {R.}~\bibnamefont
  {Gonz{\'a}lez-Hern{\'a}ndez}}, \bibinfo {author} {\bibfnamefont
  {X.}~\bibnamefont {Wang}}, \bibinfo {author} {\bibfnamefont {H.}~\bibnamefont
  {Yan}}, \bibinfo {author} {\bibfnamefont {P.}~\bibnamefont {Qin}}, \emph
  {et~al.},\ }\bibfield  {title} {\bibinfo {title} {An anomalous {H}all effect
  in altermagnetic ruthenium dioxide},\ }\href@noop {} {\bibfield  {journal}
  {\bibinfo  {journal} {Nat. Electron.}\ ,\ \bibinfo {pages} {735–743}}
  (\bibinfo {year} {2022})}\BibitemShut {NoStop}%
\bibitem [{\citenamefont {{Bychkov}}\ and\ \citenamefont
  {{Rashba}}(1984)}]{1984JETPL}%
  \BibitemOpen
  \bibfield  {author} {\bibinfo {author} {\bibfnamefont {Y.~A.}\ \bibnamefont
  {{Bychkov}}}\ and\ \bibinfo {author} {\bibfnamefont {{\'E}.~I.}\ \bibnamefont
  {{Rashba}}},\ }\bibfield  {title} {\bibinfo {title} {{Properties of a 2D
  electron gas with lifted spectral degeneracy}},\ }\href@noop {} {\bibfield
  {journal} {\bibinfo  {journal} {JETP Lett.}\ }\textbf {\bibinfo {volume}
  {39}},\ \bibinfo {pages} {78} (\bibinfo {year} {1984})}\BibitemShut {NoStop}%
\bibitem [{\citenamefont {Manchon}\ \emph {et~al.}(2015)\citenamefont
  {Manchon}, \citenamefont {Koo}, \citenamefont {Nitta}, \citenamefont
  {Frolov},\ and\ \citenamefont {Duine}}]{manchon2015new}%
  \BibitemOpen
  \bibfield  {author} {\bibinfo {author} {\bibfnamefont {A.}~\bibnamefont
  {Manchon}}, \bibinfo {author} {\bibfnamefont {H.~C.}\ \bibnamefont {Koo}},
  \bibinfo {author} {\bibfnamefont {J.}~\bibnamefont {Nitta}}, \bibinfo
  {author} {\bibfnamefont {S.}~\bibnamefont {Frolov}},\ and\ \bibinfo {author}
  {\bibfnamefont {R.}~\bibnamefont {Duine}},\ }\bibfield  {title} {\bibinfo
  {title} {New perspectives for {R}ashba spin--orbit coupling},\ }\href@noop {}
  {\bibfield  {journal} {\bibinfo  {journal} {Nat. Mater.}\ }\textbf {\bibinfo
  {volume} {14}},\ \bibinfo {pages} {871} (\bibinfo {year} {2015})}\BibitemShut
  {NoStop}%
\bibitem [{\citenamefont {Avci}\ \emph {et~al.}(2015)\citenamefont {Avci},
  \citenamefont {Garello}, \citenamefont {Ghosh}, \citenamefont {Gabureac},
  \citenamefont {Alvarado},\ and\ \citenamefont
  {Gambardella}}]{avci2015unidirectional}%
  \BibitemOpen
  \bibfield  {author} {\bibinfo {author} {\bibfnamefont {C.~O.}\ \bibnamefont
  {Avci}}, \bibinfo {author} {\bibfnamefont {K.}~\bibnamefont {Garello}},
  \bibinfo {author} {\bibfnamefont {A.}~\bibnamefont {Ghosh}}, \bibinfo
  {author} {\bibfnamefont {M.}~\bibnamefont {Gabureac}}, \bibinfo {author}
  {\bibfnamefont {S.~F.}\ \bibnamefont {Alvarado}},\ and\ \bibinfo {author}
  {\bibfnamefont {P.}~\bibnamefont {Gambardella}},\ }\bibfield  {title}
  {\bibinfo {title} {Unidirectional spin {H}all magnetoresistance in
  ferromagnet/normal metal bilayers},\ }\href@noop {} {\bibfield  {journal}
  {\bibinfo  {journal} {Nat. Phys.}\ }\textbf {\bibinfo {volume} {11}},\
  \bibinfo {pages} {570} (\bibinfo {year} {2015})}\BibitemShut {NoStop}%
\bibitem [{\citenamefont {Fiebig}\ \emph {et~al.}(2016)\citenamefont {Fiebig},
  \citenamefont {Lottermoser}, \citenamefont {Meier},\ and\ \citenamefont
  {Trassin}}]{fiebig2016evolution}%
  \BibitemOpen
  \bibfield  {author} {\bibinfo {author} {\bibfnamefont {M.}~\bibnamefont
  {Fiebig}}, \bibinfo {author} {\bibfnamefont {T.}~\bibnamefont {Lottermoser}},
  \bibinfo {author} {\bibfnamefont {D.}~\bibnamefont {Meier}},\ and\ \bibinfo
  {author} {\bibfnamefont {M.}~\bibnamefont {Trassin}},\ }\bibfield  {title}
  {\bibinfo {title} {The evolution of multiferroics},\ }\href@noop {}
  {\bibfield  {journal} {\bibinfo  {journal} {Nat. Rev. Mater.}\ }\textbf
  {\bibinfo {volume} {1}},\ \bibinfo {pages} {1} (\bibinfo {year}
  {2016})}\BibitemShut {NoStop}%
\bibitem [{\citenamefont {Ideue}\ \emph {et~al.}(2017)\citenamefont {Ideue},
  \citenamefont {Hamamoto}, \citenamefont {Koshikawa}, \citenamefont {Ezawa},
  \citenamefont {Shimizu}, \citenamefont {Kaneko}, \citenamefont {Tokura},
  \citenamefont {Nagaosa},\ and\ \citenamefont {Iwasa}}]{ideue2017bulk}%
  \BibitemOpen
  \bibfield  {author} {\bibinfo {author} {\bibfnamefont {T.}~\bibnamefont
  {Ideue}}, \bibinfo {author} {\bibfnamefont {K.}~\bibnamefont {Hamamoto}},
  \bibinfo {author} {\bibfnamefont {S.}~\bibnamefont {Koshikawa}}, \bibinfo
  {author} {\bibfnamefont {M.}~\bibnamefont {Ezawa}}, \bibinfo {author}
  {\bibfnamefont {S.}~\bibnamefont {Shimizu}}, \bibinfo {author} {\bibfnamefont
  {Y.}~\bibnamefont {Kaneko}}, \bibinfo {author} {\bibfnamefont
  {Y.}~\bibnamefont {Tokura}}, \bibinfo {author} {\bibfnamefont
  {N.}~\bibnamefont {Nagaosa}},\ and\ \bibinfo {author} {\bibfnamefont
  {Y.}~\bibnamefont {Iwasa}},\ }\bibfield  {title} {\bibinfo {title} {Bulk
  rectification effect in a polar semiconductor},\ }\href@noop {} {\bibfield
  {journal} {\bibinfo  {journal} {Nat. Phys.}\ }\textbf {\bibinfo {volume}
  {13}},\ \bibinfo {pages} {578} (\bibinfo {year} {2017})}\BibitemShut
  {NoStop}%
\bibitem [{\citenamefont {Br{\'e}hin}\ \emph {et~al.}(2023)\citenamefont
  {Br{\'e}hin}, \citenamefont {Chen}, \citenamefont {D’Antuono},
  \citenamefont {Varotto}, \citenamefont {Stornaiuolo}, \citenamefont
  {Piamonteze}, \citenamefont {Varignon}, \citenamefont {Salluzzo},\ and\
  \citenamefont {Bibes}}]{brehin2023coexistence}%
  \BibitemOpen
  \bibfield  {author} {\bibinfo {author} {\bibfnamefont {J.}~\bibnamefont
  {Br{\'e}hin}}, \bibinfo {author} {\bibfnamefont {Y.}~\bibnamefont {Chen}},
  \bibinfo {author} {\bibfnamefont {M.}~\bibnamefont {D’Antuono}}, \bibinfo
  {author} {\bibfnamefont {S.}~\bibnamefont {Varotto}}, \bibinfo {author}
  {\bibfnamefont {D.}~\bibnamefont {Stornaiuolo}}, \bibinfo {author}
  {\bibfnamefont {C.}~\bibnamefont {Piamonteze}}, \bibinfo {author}
  {\bibfnamefont {J.}~\bibnamefont {Varignon}}, \bibinfo {author}
  {\bibfnamefont {M.}~\bibnamefont {Salluzzo}},\ and\ \bibinfo {author}
  {\bibfnamefont {M.}~\bibnamefont {Bibes}},\ }\bibfield  {title} {\bibinfo
  {title} {Coexistence and coupling of ferroelectricity and magnetism in an
  oxide two-dimensional electron gas},\ }\href@noop {} {\bibfield  {journal}
  {\bibinfo  {journal} {Nat. Phys.}\ }\textbf {\bibinfo {volume} {19}},\
  \bibinfo {pages} {823} (\bibinfo {year} {2023})}\BibitemShut {NoStop}%
\bibitem [{\citenamefont {Landau}\ \emph {et~al.}(1995)\citenamefont {Landau},
  \citenamefont {Lifshitz},\ and\ \citenamefont
  {Pitaevskii}}]{landau1995electrodynamics}%
  \BibitemOpen
  \bibfield  {author} {\bibinfo {author} {\bibfnamefont {L.}~\bibnamefont
  {Landau}}, \bibinfo {author} {\bibfnamefont {E.}~\bibnamefont {Lifshitz}},\
  and\ \bibinfo {author} {\bibfnamefont {L.}~\bibnamefont {Pitaevskii}},\
  }\href {https://books.google.de/books?id=OZrFzwEACAAJ} {\emph {\bibinfo
  {title} {Electrodynamics of Continuous Media: Volume 8}}},\ Course of
  theoretical physics\ (\bibinfo  {publisher} {Elsevier Science},\ \bibinfo
  {year} {1995})\BibitemShut {NoStop}%
\bibitem [{\citenamefont {Sunko}\ \emph {et~al.}(2017)\citenamefont {Sunko},
  \citenamefont {Rosner}, \citenamefont {Kushwaha}, \citenamefont {Khim},
  \citenamefont {Mazzola}, \citenamefont {Bawden}, \citenamefont {Clark},
  \citenamefont {Riley}, \citenamefont {Kasinathan}, \citenamefont {Haverkort}
  \emph {et~al.}}]{sunko2017maximal}%
  \BibitemOpen
  \bibfield  {author} {\bibinfo {author} {\bibfnamefont {V.}~\bibnamefont
  {Sunko}}, \bibinfo {author} {\bibfnamefont {H.}~\bibnamefont {Rosner}},
  \bibinfo {author} {\bibfnamefont {P.}~\bibnamefont {Kushwaha}}, \bibinfo
  {author} {\bibfnamefont {S.}~\bibnamefont {Khim}}, \bibinfo {author}
  {\bibfnamefont {F.}~\bibnamefont {Mazzola}}, \bibinfo {author} {\bibfnamefont
  {L.}~\bibnamefont {Bawden}}, \bibinfo {author} {\bibfnamefont
  {O.}~\bibnamefont {Clark}}, \bibinfo {author} {\bibfnamefont
  {J.}~\bibnamefont {Riley}}, \bibinfo {author} {\bibfnamefont
  {D.}~\bibnamefont {Kasinathan}}, \bibinfo {author} {\bibfnamefont
  {M.}~\bibnamefont {Haverkort}}, \emph {et~al.},\ }\bibfield  {title}
  {\bibinfo {title} {Maximal {R}ashba-like spin splitting via
  kinetic-energy-coupled inversion-symmetry breaking},\ }\href@noop {}
  {\bibfield  {journal} {\bibinfo  {journal} {Nature}\ }\textbf {\bibinfo
  {volume} {549}},\ \bibinfo {pages} {492} (\bibinfo {year}
  {2017})}\BibitemShut {NoStop}%
\bibitem [{\citenamefont {Seki}\ \emph {et~al.}(2008)\citenamefont {Seki},
  \citenamefont {Onose},\ and\ \citenamefont
  {Tokura}}]{PhysRevLett.101.067204}%
  \BibitemOpen
  \bibfield  {author} {\bibinfo {author} {\bibfnamefont {S.}~\bibnamefont
  {Seki}}, \bibinfo {author} {\bibfnamefont {Y.}~\bibnamefont {Onose}},\ and\
  \bibinfo {author} {\bibfnamefont {Y.}~\bibnamefont {Tokura}},\ }\bibfield
  {title} {\bibinfo {title} {Spin-driven ferroelectricity in triangular lattice
  antiferromagnets ${A}${C}r{O}$_{2}$ (${A}=\mathrm{Cu}$, {A}g, {L}i, or
  {N}a)},\ }\href {https://doi.org/10.1103/PhysRevLett.101.067204} {\bibfield
  {journal} {\bibinfo  {journal} {Phys. Rev. Lett.}\ }\textbf {\bibinfo
  {volume} {101}},\ \bibinfo {pages} {067204} (\bibinfo {year}
  {2008})}\BibitemShut {NoStop}%
\bibitem [{\citenamefont {Singh}\ \emph {et~al.}(2009)\citenamefont {Singh},
  \citenamefont {Maignan}, \citenamefont {Martin},\ and\ \citenamefont
  {Simon}}]{Singh2009}%
  \BibitemOpen
  \bibfield  {author} {\bibinfo {author} {\bibfnamefont {K.}~\bibnamefont
  {Singh}}, \bibinfo {author} {\bibfnamefont {A.}~\bibnamefont {Maignan}},
  \bibinfo {author} {\bibfnamefont {C.}~\bibnamefont {Martin}},\ and\ \bibinfo
  {author} {\bibfnamefont {C.}~\bibnamefont {Simon}},\ }\bibfield  {title}
  {\bibinfo {title} {{A}g{C}r{S}$_2$: A spin driven ferroelectric},\ }\href
  {https://doi.org/10.1021/cm902524h} {\bibfield  {journal} {\bibinfo
  {journal} {Chem. Mater.}\ }\textbf {\bibinfo {volume} {21}},\ \bibinfo
  {pages} {5007} (\bibinfo {year} {2009})}\BibitemShut {NoStop}%
\bibitem [{\citenamefont {Xu}\ \emph {et~al.}(2022)\citenamefont {Xu},
  \citenamefont {Zhong}, \citenamefont {Zuo}, \citenamefont {Li}, \citenamefont
  {Li}, \citenamefont {Pi}, \citenamefont {Chen}, \citenamefont {Wu},
  \citenamefont {Zhai},\ and\ \citenamefont {Zhou}}]{xiang2022}%
  \BibitemOpen
  \bibfield  {author} {\bibinfo {author} {\bibfnamefont {X.}~\bibnamefont
  {Xu}}, \bibinfo {author} {\bibfnamefont {T.}~\bibnamefont {Zhong}}, \bibinfo
  {author} {\bibfnamefont {N.}~\bibnamefont {Zuo}}, \bibinfo {author}
  {\bibfnamefont {Z.}~\bibnamefont {Li}}, \bibinfo {author} {\bibfnamefont
  {D.}~\bibnamefont {Li}}, \bibinfo {author} {\bibfnamefont {L.}~\bibnamefont
  {Pi}}, \bibinfo {author} {\bibfnamefont {P.}~\bibnamefont {Chen}}, \bibinfo
  {author} {\bibfnamefont {M.}~\bibnamefont {Wu}}, \bibinfo {author}
  {\bibfnamefont {T.}~\bibnamefont {Zhai}},\ and\ \bibinfo {author}
  {\bibfnamefont {X.}~\bibnamefont {Zhou}},\ }\bibfield  {title} {\bibinfo
  {title} {High-{T}c two-dimensional ferroelectric {C}u{C}r{S}$_2$ grown via
  chemical vapor deposition},\ }\href {https://doi.org/10.1021/acsnano.2c01470}
  {\bibfield  {journal} {\bibinfo  {journal} {ACS Nano}\ }\textbf {\bibinfo
  {volume} {16}},\ \bibinfo {pages} {8141} (\bibinfo {year}
  {2022})}\BibitemShut {NoStop}%
\bibitem [{\citenamefont {Gascoin}\ and\ \citenamefont
  {Maignan}(2011)}]{gascoin2011order}%
  \BibitemOpen
  \bibfield  {author} {\bibinfo {author} {\bibfnamefont {F.}~\bibnamefont
  {Gascoin}}\ and\ \bibinfo {author} {\bibfnamefont {A.}~\bibnamefont
  {Maignan}},\ }\bibfield  {title} {\bibinfo {title} {Order--disorder
  transition in {A}g{C}r{S}e\textsubscript{2}: A new route to efficient
  thermoelectrics},\ }\href@noop {} {\bibfield  {journal} {\bibinfo  {journal}
  {Chem. Mater.}\ }\textbf {\bibinfo {volume} {23}},\ \bibinfo {pages} {2510}
  (\bibinfo {year} {2011})}\BibitemShut {NoStop}%
\bibitem [{\citenamefont {Li}\ \emph {et~al.}(2018)\citenamefont {Li},
  \citenamefont {Wang}, \citenamefont {Kawakita}, \citenamefont {Zhang},
  \citenamefont {Feygenson}, \citenamefont {Yu}, \citenamefont {Wu},
  \citenamefont {Ohara}, \citenamefont {Kikuchi}, \citenamefont {Shibata} \emph
  {et~al.}}]{li2018liquid}%
  \BibitemOpen
  \bibfield  {author} {\bibinfo {author} {\bibfnamefont {B.}~\bibnamefont
  {Li}}, \bibinfo {author} {\bibfnamefont {H.}~\bibnamefont {Wang}}, \bibinfo
  {author} {\bibfnamefont {Y.}~\bibnamefont {Kawakita}}, \bibinfo {author}
  {\bibfnamefont {Q.}~\bibnamefont {Zhang}}, \bibinfo {author} {\bibfnamefont
  {M.}~\bibnamefont {Feygenson}}, \bibinfo {author} {\bibfnamefont
  {H.}~\bibnamefont {Yu}}, \bibinfo {author} {\bibfnamefont {D.}~\bibnamefont
  {Wu}}, \bibinfo {author} {\bibfnamefont {K.}~\bibnamefont {Ohara}}, \bibinfo
  {author} {\bibfnamefont {T.}~\bibnamefont {Kikuchi}}, \bibinfo {author}
  {\bibfnamefont {K.}~\bibnamefont {Shibata}}, \emph {et~al.},\ }\bibfield
  {title} {\bibinfo {title} {Liquid-like thermal conduction in intercalated
  layered crystalline solids},\ }\href@noop {} {\bibfield  {journal} {\bibinfo
  {journal} {Nat. Mater.}\ }\textbf {\bibinfo {volume} {17}},\ \bibinfo {pages}
  {226} (\bibinfo {year} {2018})}\BibitemShut {NoStop}%
\bibitem [{\citenamefont {Ding}\ \emph {et~al.}(2020)\citenamefont {Ding},
  \citenamefont {Niedziela}, \citenamefont {Bansal}, \citenamefont {Wang},
  \citenamefont {He}, \citenamefont {May}, \citenamefont {Ehlers},
  \citenamefont {Abernathy}, \citenamefont {Said}, \citenamefont {Alatas},
  \citenamefont {Ren}, \citenamefont {Arya},\ and\ \citenamefont
  {Delaire}}]{PNASDing2020}%
  \BibitemOpen
  \bibfield  {author} {\bibinfo {author} {\bibfnamefont {J.}~\bibnamefont
  {Ding}}, \bibinfo {author} {\bibfnamefont {J.~L.}\ \bibnamefont {Niedziela}},
  \bibinfo {author} {\bibfnamefont {D.}~\bibnamefont {Bansal}}, \bibinfo
  {author} {\bibfnamefont {J.}~\bibnamefont {Wang}}, \bibinfo {author}
  {\bibfnamefont {X.}~\bibnamefont {He}}, \bibinfo {author} {\bibfnamefont
  {A.~F.}\ \bibnamefont {May}}, \bibinfo {author} {\bibfnamefont
  {G.}~\bibnamefont {Ehlers}}, \bibinfo {author} {\bibfnamefont {D.~L.}\
  \bibnamefont {Abernathy}}, \bibinfo {author} {\bibfnamefont {A.}~\bibnamefont
  {Said}}, \bibinfo {author} {\bibfnamefont {A.}~\bibnamefont {Alatas}},
  \bibinfo {author} {\bibfnamefont {Y.}~\bibnamefont {Ren}}, \bibinfo {author}
  {\bibfnamefont {G.}~\bibnamefont {Arya}},\ and\ \bibinfo {author}
  {\bibfnamefont {O.}~\bibnamefont {Delaire}},\ }\bibfield  {title} {\bibinfo
  {title} {Anharmonic lattice dynamics and superionic transition in
  {A}g{C}r{S}e$_2$},\ }\href {https://doi.org/10.1073/pnas.1913916117}
  {\bibfield  {journal} {\bibinfo  {journal} {Proc. Natl. Acad. Sci. U.S.A.}\
  }\textbf {\bibinfo {volume} {117}},\ \bibinfo {pages} {3930} (\bibinfo {year}
  {2020})}\BibitemShut {NoStop}%
\bibitem [{\citenamefont {Baenitz}\ \emph {et~al.}(2021)\citenamefont
  {Baenitz}, \citenamefont {Piva}, \citenamefont {Luther}, \citenamefont
  {Sichelschmidt}, \citenamefont {Ranjith}, \citenamefont {Dawczak-Debicki},
  \citenamefont {Ajeesh}, \citenamefont {Kim}, \citenamefont {Siemann},
  \citenamefont {Bigi}, \citenamefont {Manuel}, \citenamefont {Khalyavin},
  \citenamefont {Sokolov}, \citenamefont {Mokhtari}, \citenamefont {Zhang},
  \citenamefont {Yasuoka}, \citenamefont {King}, \citenamefont {Vinai},
  \citenamefont {Polewczyk}, \citenamefont {Torelli}, \citenamefont {Wosnitza},
  \citenamefont {Burkhardt}, \citenamefont {Schmidt}, \citenamefont {Rosner},
  \citenamefont {Wirth}, \citenamefont {K\"uhne}, \citenamefont {Nicklas},\
  and\ \citenamefont {Schmidt}}]{Baenitz}%
  \BibitemOpen
  \bibfield  {author} {\bibinfo {author} {\bibfnamefont {M.}~\bibnamefont
  {Baenitz}}, \bibinfo {author} {\bibfnamefont {M.~M.}\ \bibnamefont {Piva}},
  \bibinfo {author} {\bibfnamefont {S.}~\bibnamefont {Luther}}, \bibinfo
  {author} {\bibfnamefont {J.}~\bibnamefont {Sichelschmidt}}, \bibinfo {author}
  {\bibfnamefont {K.~M.}\ \bibnamefont {Ranjith}}, \bibinfo {author}
  {\bibfnamefont {H.}~\bibnamefont {Dawczak-Debicki}}, \bibinfo {author}
  {\bibfnamefont {M.~O.}\ \bibnamefont {Ajeesh}}, \bibinfo {author}
  {\bibfnamefont {S.-J.}\ \bibnamefont {Kim}}, \bibinfo {author} {\bibfnamefont
  {G.}~\bibnamefont {Siemann}}, \bibinfo {author} {\bibfnamefont
  {C.}~\bibnamefont {Bigi}}, \bibinfo {author} {\bibfnamefont {P.}~\bibnamefont
  {Manuel}}, \bibinfo {author} {\bibfnamefont {D.}~\bibnamefont {Khalyavin}},
  \bibinfo {author} {\bibfnamefont {D.~A.}\ \bibnamefont {Sokolov}}, \bibinfo
  {author} {\bibfnamefont {P.}~\bibnamefont {Mokhtari}}, \bibinfo {author}
  {\bibfnamefont {H.}~\bibnamefont {Zhang}}, \bibinfo {author} {\bibfnamefont
  {H.}~\bibnamefont {Yasuoka}}, \bibinfo {author} {\bibfnamefont {P.~D.~C.}\
  \bibnamefont {King}}, \bibinfo {author} {\bibfnamefont {G.}~\bibnamefont
  {Vinai}}, \bibinfo {author} {\bibfnamefont {V.}~\bibnamefont {Polewczyk}},
  \bibinfo {author} {\bibfnamefont {P.}~\bibnamefont {Torelli}}, \bibinfo
  {author} {\bibfnamefont {J.}~\bibnamefont {Wosnitza}}, \bibinfo {author}
  {\bibfnamefont {U.}~\bibnamefont {Burkhardt}}, \bibinfo {author}
  {\bibfnamefont {B.}~\bibnamefont {Schmidt}}, \bibinfo {author} {\bibfnamefont
  {H.}~\bibnamefont {Rosner}}, \bibinfo {author} {\bibfnamefont
  {S.}~\bibnamefont {Wirth}}, \bibinfo {author} {\bibfnamefont
  {H.}~\bibnamefont {K\"uhne}}, \bibinfo {author} {\bibfnamefont
  {M.}~\bibnamefont {Nicklas}},\ and\ \bibinfo {author} {\bibfnamefont
  {M.}~\bibnamefont {Schmidt}},\ }\bibfield  {title} {\bibinfo {title} {Planar
  triangular ${S}=3/2$ magnet {A}g{C}r{S}e$_{2}$: Magnetic frustration, short
  range correlations, and field-tuned anisotropic cycloidal magnetic order},\
  }\href {https://doi.org/10.1103/PhysRevB.104.134410} {\bibfield  {journal}
  {\bibinfo  {journal} {Phys. Rev. B}\ }\textbf {\bibinfo {volume} {104}},\
  \bibinfo {pages} {134410} (\bibinfo {year} {2021})}\BibitemShut {NoStop}%
\bibitem [{\citenamefont {Takahashi}\ \emph {et~al.}(2022)\citenamefont
  {Takahashi}, \citenamefont {Akiba}, \citenamefont {Mayo}, \citenamefont
  {Akiba}, \citenamefont {Miyake}, \citenamefont {Tokunaga}, \citenamefont
  {Mori}, \citenamefont {Arita},\ and\ \citenamefont
  {Ishiwata}}]{PhysRevMaterials.6.054602}%
  \BibitemOpen
  \bibfield  {author} {\bibinfo {author} {\bibfnamefont {H.}~\bibnamefont
  {Takahashi}}, \bibinfo {author} {\bibfnamefont {T.}~\bibnamefont {Akiba}},
  \bibinfo {author} {\bibfnamefont {A.~H.}\ \bibnamefont {Mayo}}, \bibinfo
  {author} {\bibfnamefont {K.}~\bibnamefont {Akiba}}, \bibinfo {author}
  {\bibfnamefont {A.}~\bibnamefont {Miyake}}, \bibinfo {author} {\bibfnamefont
  {M.}~\bibnamefont {Tokunaga}}, \bibinfo {author} {\bibfnamefont
  {H.}~\bibnamefont {Mori}}, \bibinfo {author} {\bibfnamefont {R.}~\bibnamefont
  {Arita}},\ and\ \bibinfo {author} {\bibfnamefont {S.}~\bibnamefont
  {Ishiwata}},\ }\bibfield  {title} {\bibinfo {title} {Spin-orbit-derived giant
  magnetoresistance in a layered magnetic semiconductor {A}g{C}r{S}e$_2$},\
  }\href {https://doi.org/10.1103/PhysRevMaterials.6.054602} {\bibfield
  {journal} {\bibinfo  {journal} {Phys. Rev. Mater.}\ }\textbf {\bibinfo
  {volume} {6}},\ \bibinfo {pages} {054602} (\bibinfo {year}
  {2022})}\BibitemShut {NoStop}%
\bibitem [{\citenamefont {Engelsman}\ \emph {et~al.}(1973)\citenamefont
  {Engelsman}, \citenamefont {Wiegers}, \citenamefont {Jellinek},\ and\
  \citenamefont {Van~Laar}}]{engelsman1973crystal}%
  \BibitemOpen
  \bibfield  {author} {\bibinfo {author} {\bibfnamefont {F.}~\bibnamefont
  {Engelsman}}, \bibinfo {author} {\bibfnamefont {G.}~\bibnamefont {Wiegers}},
  \bibinfo {author} {\bibfnamefont {F.}~\bibnamefont {Jellinek}},\ and\
  \bibinfo {author} {\bibfnamefont {B.}~\bibnamefont {Van~Laar}},\ }\bibfield
  {title} {\bibinfo {title} {Crystal structures and magnetic structures of some
  metal (i) chromium (iii) sulfides and selenides},\ }\href@noop {} {\bibfield
  {journal} {\bibinfo  {journal} {J. Solid State Chem.}\ }\textbf {\bibinfo
  {volume} {6}},\ \bibinfo {pages} {574} (\bibinfo {year} {1973})}\BibitemShut
  {NoStop}%
\bibitem [{\citenamefont {Damay}\ \emph {et~al.}(2016)\citenamefont {Damay},
  \citenamefont {Petit}, \citenamefont {Rols}, \citenamefont {Braendlein},
  \citenamefont {Daou}, \citenamefont {Elka{\"\i}m}, \citenamefont {Fauth},
  \citenamefont {Gascoin}, \citenamefont {Martin},\ and\ \citenamefont
  {Maignan}}]{damay2016localised}%
  \BibitemOpen
  \bibfield  {author} {\bibinfo {author} {\bibfnamefont {F.}~\bibnamefont
  {Damay}}, \bibinfo {author} {\bibfnamefont {S.}~\bibnamefont {Petit}},
  \bibinfo {author} {\bibfnamefont {S.}~\bibnamefont {Rols}}, \bibinfo {author}
  {\bibfnamefont {M.}~\bibnamefont {Braendlein}}, \bibinfo {author}
  {\bibfnamefont {R.}~\bibnamefont {Daou}}, \bibinfo {author} {\bibfnamefont
  {E.}~\bibnamefont {Elka{\"\i}m}}, \bibinfo {author} {\bibfnamefont
  {F.}~\bibnamefont {Fauth}}, \bibinfo {author} {\bibfnamefont
  {F.}~\bibnamefont {Gascoin}}, \bibinfo {author} {\bibfnamefont
  {C.}~\bibnamefont {Martin}},\ and\ \bibinfo {author} {\bibfnamefont
  {A.}~\bibnamefont {Maignan}},\ }\bibfield  {title} {\bibinfo {title}
  {Localised {A}g$^+$ vibrations at the origin of ultralow thermal conductivity
  in layered thermoelectric {A}g{C}r{S}e\textsubscript{2}},\ }\href@noop {}
  {\bibfield  {journal} {\bibinfo  {journal} {Sci. Rep.}\ }\textbf {\bibinfo
  {volume} {6}},\ \bibinfo {pages} {1} (\bibinfo {year} {2016})}\BibitemShut
  {NoStop}%
\bibitem [{\citenamefont {Nakatsuji}\ \emph {et~al.}(2015)\citenamefont
  {Nakatsuji}, \citenamefont {Kiyohara},\ and\ \citenamefont
  {Higo}}]{nakatsuji2015large}%
  \BibitemOpen
  \bibfield  {author} {\bibinfo {author} {\bibfnamefont {S.}~\bibnamefont
  {Nakatsuji}}, \bibinfo {author} {\bibfnamefont {N.}~\bibnamefont
  {Kiyohara}},\ and\ \bibinfo {author} {\bibfnamefont {T.}~\bibnamefont
  {Higo}},\ }\bibfield  {title} {\bibinfo {title} {Large anomalous {H}all
  effect in a non-collinear antiferromagnet at room temperature},\ }\href@noop
  {} {\bibfield  {journal} {\bibinfo  {journal} {Nature}\ }\textbf {\bibinfo
  {volume} {527}},\ \bibinfo {pages} {212} (\bibinfo {year}
  {2015})}\BibitemShut {NoStop}%
\bibitem [{\citenamefont {Nayak}\ \emph {et~al.}(2016)\citenamefont {Nayak},
  \citenamefont {Fischer}, \citenamefont {Sun}, \citenamefont {Yan},
  \citenamefont {Karel}, \citenamefont {Komarek}, \citenamefont {Shekhar},
  \citenamefont {Kumar}, \citenamefont {Schnelle}, \citenamefont {K{\"u}bler}
  \emph {et~al.}}]{nayak2016large}%
  \BibitemOpen
  \bibfield  {author} {\bibinfo {author} {\bibfnamefont {A.~K.}\ \bibnamefont
  {Nayak}}, \bibinfo {author} {\bibfnamefont {J.~E.}\ \bibnamefont {Fischer}},
  \bibinfo {author} {\bibfnamefont {Y.}~\bibnamefont {Sun}}, \bibinfo {author}
  {\bibfnamefont {B.}~\bibnamefont {Yan}}, \bibinfo {author} {\bibfnamefont
  {J.}~\bibnamefont {Karel}}, \bibinfo {author} {\bibfnamefont {A.~C.}\
  \bibnamefont {Komarek}}, \bibinfo {author} {\bibfnamefont {C.}~\bibnamefont
  {Shekhar}}, \bibinfo {author} {\bibfnamefont {N.}~\bibnamefont {Kumar}},
  \bibinfo {author} {\bibfnamefont {W.}~\bibnamefont {Schnelle}}, \bibinfo
  {author} {\bibfnamefont {J.}~\bibnamefont {K{\"u}bler}}, \emph {et~al.},\
  }\bibfield  {title} {\bibinfo {title} {Large anomalous {H}all effect driven
  by a nonvanishing {B}erry curvature in the noncolinear antiferromagnet
  {M}n\textsubscript{3}{G}e},\ }\href@noop {} {\bibfield  {journal} {\bibinfo
  {journal} {Sci. Adv.}\ }\textbf {\bibinfo {volume} {2}},\ \bibinfo {pages}
  {e1501870} (\bibinfo {year} {2016})}\BibitemShut {NoStop}%
\bibitem [{\citenamefont {Ghimire}\ \emph {et~al.}(2018)\citenamefont
  {Ghimire}, \citenamefont {Botana}, \citenamefont {Jiang}, \citenamefont
  {Zhang}, \citenamefont {Chen},\ and\ \citenamefont
  {Mitchell}}]{ghimire2018large}%
  \BibitemOpen
  \bibfield  {author} {\bibinfo {author} {\bibfnamefont {N.~J.}\ \bibnamefont
  {Ghimire}}, \bibinfo {author} {\bibfnamefont {A.}~\bibnamefont {Botana}},
  \bibinfo {author} {\bibfnamefont {J.}~\bibnamefont {Jiang}}, \bibinfo
  {author} {\bibfnamefont {J.}~\bibnamefont {Zhang}}, \bibinfo {author}
  {\bibfnamefont {Y.-S.}\ \bibnamefont {Chen}},\ and\ \bibinfo {author}
  {\bibfnamefont {J.}~\bibnamefont {Mitchell}},\ }\bibfield  {title} {\bibinfo
  {title} {Large anomalous {H}all effect in the chiral-lattice antiferromagnet
  {C}o{N}b\textsubscript{3}{S}\textsubscript{6}},\ }\href@noop {} {\bibfield
  {journal} {\bibinfo  {journal} {Nat. Commun.}\ }\textbf {\bibinfo {volume}
  {9}},\ \bibinfo {pages} {1} (\bibinfo {year} {2018})}\BibitemShut {NoStop}%
\bibitem [{\citenamefont {Lee}\ \emph {et~al.}(2019)\citenamefont {Lee},
  \citenamefont {Zhu}, \citenamefont {Wang}, \citenamefont {Miao},
  \citenamefont {Pillsbury}, \citenamefont {Yi}, \citenamefont {Kempinger},
  \citenamefont {Hu}, \citenamefont {Heikes}, \citenamefont {Quarterman} \emph
  {et~al.}}]{lee2019spin}%
  \BibitemOpen
  \bibfield  {author} {\bibinfo {author} {\bibfnamefont {S.~H.}\ \bibnamefont
  {Lee}}, \bibinfo {author} {\bibfnamefont {Y.}~\bibnamefont {Zhu}}, \bibinfo
  {author} {\bibfnamefont {Y.}~\bibnamefont {Wang}}, \bibinfo {author}
  {\bibfnamefont {L.}~\bibnamefont {Miao}}, \bibinfo {author} {\bibfnamefont
  {T.}~\bibnamefont {Pillsbury}}, \bibinfo {author} {\bibfnamefont
  {H.}~\bibnamefont {Yi}}, \bibinfo {author} {\bibfnamefont {S.}~\bibnamefont
  {Kempinger}}, \bibinfo {author} {\bibfnamefont {J.}~\bibnamefont {Hu}},
  \bibinfo {author} {\bibfnamefont {C.~A.}\ \bibnamefont {Heikes}}, \bibinfo
  {author} {\bibfnamefont {P.}~\bibnamefont {Quarterman}}, \emph {et~al.},\
  }\bibfield  {title} {\bibinfo {title} {Spin scattering and noncollinear spin
  structure-induced intrinsic anomalous {H}all effect in antiferromagnetic
  topological insulator {M}n{B}i\textsubscript{2}{T}e\textsubscript{4}},\
  }\href@noop {} {\bibfield  {journal} {\bibinfo  {journal} {Phys. Rev. Res.}\
  }\textbf {\bibinfo {volume} {1}},\ \bibinfo {pages} {012011} (\bibinfo {year}
  {2019})}\BibitemShut {NoStop}%
\bibitem [{\citenamefont {Dzyaloshinsky}(1958)}]{DZYALOSHINSKY1958241}%
  \BibitemOpen
  \bibfield  {author} {\bibinfo {author} {\bibfnamefont {I.}~\bibnamefont
  {Dzyaloshinsky}},\ }\bibfield  {title} {\bibinfo {title} {A thermodynamic
  theory of “weak” ferromagnetism of antiferromagnetics},\ }\href
  {https://doi.org/https://doi.org/10.1016/0022-3697(58)90076-3} {\bibfield
  {journal} {\bibinfo  {journal} {Journal of Physics and Chemistry of Solids}\
  }\textbf {\bibinfo {volume} {4}},\ \bibinfo {pages} {241} (\bibinfo {year}
  {1958})}\BibitemShut {NoStop}%
\bibitem [{\citenamefont {Moriya}(1960)}]{PhysRev.120.91}%
  \BibitemOpen
  \bibfield  {author} {\bibinfo {author} {\bibfnamefont {T.}~\bibnamefont
  {Moriya}},\ }\bibfield  {title} {\bibinfo {title} {Anisotropic superexchange
  interaction and weak ferromagnetism},\ }\href
  {https://doi.org/10.1103/PhysRev.120.91} {\bibfield  {journal} {\bibinfo
  {journal} {Phys. Rev.}\ }\textbf {\bibinfo {volume} {120}},\ \bibinfo {pages}
  {91} (\bibinfo {year} {1960})}\BibitemShut {NoStop}%
\bibitem [{\citenamefont {Liu}\ \emph {et~al.}(2013)\citenamefont {Liu},
  \citenamefont {Hsu},\ and\ \citenamefont {Liu}}]{PhysRevLett.111.086802}%
  \BibitemOpen
  \bibfield  {author} {\bibinfo {author} {\bibfnamefont {X.}~\bibnamefont
  {Liu}}, \bibinfo {author} {\bibfnamefont {H.-C.}\ \bibnamefont {Hsu}},\ and\
  \bibinfo {author} {\bibfnamefont {C.-X.}\ \bibnamefont {Liu}},\ }\bibfield
  {title} {\bibinfo {title} {In-plane magnetization-induced quantum anomalous
  hall effect},\ }\href {https://doi.org/10.1103/PhysRevLett.111.086802}
  {\bibfield  {journal} {\bibinfo  {journal} {Phys. Rev. Lett.}\ }\textbf
  {\bibinfo {volume} {111}},\ \bibinfo {pages} {086802} (\bibinfo {year}
  {2013})}\BibitemShut {NoStop}%
\bibitem [{\citenamefont {Chen}\ \emph {et~al.}(2014)\citenamefont {Chen},
  \citenamefont {Niu},\ and\ \citenamefont
  {MacDonald}}]{PhysRevLett.112.017205}%
  \BibitemOpen
  \bibfield  {author} {\bibinfo {author} {\bibfnamefont {H.}~\bibnamefont
  {Chen}}, \bibinfo {author} {\bibfnamefont {Q.}~\bibnamefont {Niu}},\ and\
  \bibinfo {author} {\bibfnamefont {A.~H.}\ \bibnamefont {MacDonald}},\
  }\bibfield  {title} {\bibinfo {title} {Anomalous {H}all effect arising from
  noncollinear antiferromagnetism},\ }\href
  {https://doi.org/10.1103/PhysRevLett.112.017205} {\bibfield  {journal}
  {\bibinfo  {journal} {Phys. Rev. Lett.}\ }\textbf {\bibinfo {volume} {112}},\
  \bibinfo {pages} {017205} (\bibinfo {year} {2014})}\BibitemShut {NoStop}%
\end{thebibliography}%

\end{document}